\documentclass[final,authoryear,12pt]{elsarticle}
\usepackage{fullpage}
\usepackage{fltpage}
\usepackage{tikz}
\usepackage[pdftex,bookmarks=true, colorlinks, citecolor=blue]{hyperref}
\usepackage{doi}
\usepackage{circuitikz}
\usepackage{graphicx}
\usepackage{todonotes}
\usepackage{lineno}
\usetikzlibrary{mindmap,backgrounds,snakes,matrix}

\newcommand{\figref}[1]{Figure~\ref{#1}}
\newcommand{\figrefp}[1]{(Figure~\ref{#1})}
\journal{Quaternary Science Reviews}
\begin{document}

\begin{frontmatter}
\title{Traditional and novel approaches to palaeoclimate modelling}
\author{Michel Crucifix} \address{ Earth and Life Institute, George Lemaitre Centre for Earth and Climate Research, Universit\'e catholique de Louvain, Louvain-la-Neuve, Belgium. E-mail:  michel.crucifix@uclouvain.be }
\newpage
\setpagewiselinenumbers
\modulolinenumbers[5]
\linenumbers
\begin{abstract}
Palaeoclimate archives contain information on climate variability, trends and mechanisms. Models are developed to explain observations and  predict  the response of the climate system to perturbations, in particular perturbations associated with  the anthropogenic influence. Here, we review three classical frameworks of climate modelling: conceptual, simulator-based (including general circulation models and Earth system models of intermediate complexity), and statistical. The conceptual framework aims at a parsimonious representation of a given climate phenomenon; the simulator-based framework connects physical and biogeochemical principles with phenomena at different spatial and temporal scales; and statistical modelling is a framework for inference from observations, given hypotheses on systematic and random effects. Recently, solutions have been proposed in the literature to combine these frameworks, and new concepts have emerged: the emulator (a statistical, computing efficient surrogate for the simulator) and the discrepancy, which is a statistical representation of the difference between the simulator and the real phenomenon. These concepts are explained, with references to implementations for both time-slices and dynamical applications. \newline
highlights:
\begin{enumerate}
\item Clarify three palaeoclimate modelling frameworks
\item Different definitions and applications of the conceptual models
\item Comment on usage and interest of numerical simulators
\item Show how statistical methods apply to process-based modelling
\item Review the current use of emulators and discrepancy in palaeoclimate modelling.
\end{enumerate}
\end{abstract}
\end{frontmatter}
\section{Introduction}
Palaeoclimatology, as any natural science, depends on observations to generate scientific activity. Questions about ice ages, Dansgaard-Oeschger events, fluctuations in CO$_2$ atmospheric concentration, would not have been raised if these phenomena had not been observed in the first place. 
Scientists attempt to model these phenomena in order to explain them, to predict them, raise new questions and suggest additional experiments. 

In climate science modellers are confronted to a dilemma about model complexity. Although this dilemma is often presented in terms of a compromise about computing power, there is a more fundamental dichotomy.

On the one hand, the explanatory power of a model tends to be better accepted if the number of \textit{ad hoc} hypotheses  is as small as possible, even if this comes at the cost of not reproducing all the details of the phenomenon. This is the celebrated Ockham's razor principle, that encourages us to explain as much as possible with as little as possible. Consistent with this line of reasoning, \cite{held05hierarchy} recommends systematic research on simple hydrodynamical models such as the dry baroclinic atmosphere. As shall be reviewed here, even simpler systems, based on a small number of equations, have non-trivial properties which are helpful to interpret glacial-interglacial cycles and abrupt events. Furthermore, small mathematical models are easy to communicate so that the results can be well reproduced by  peer scientists on the basis of information available in the publication.

On the other hand, climate scientists have developed large numerical models, which encapsulate available knowledge on a huge variety of climate processes at different spatial and temporal scales, ranging from cloud formation to sediment kinematics. Here we will term call these models: simulators.
Current climate simulators are impressively successful at reproducing many features of our climate system \footnote{a list of references is available at \url{http://www-pcmdi.llnl.gov/ipcc/subproject\_publications.php}.}, and the idea is to use them as experimental substitutes to real climate system. However, running simulators is a technically involved operation, and it is difficult if not impossible to fully appreciate the consequences of all the technical choices and physical hypotheses that they contain \citep{Winsberg99aa}.

A number of concepts and theories may help us to articulate these different modelling frameworks. 
The starting point is that mathematical models generate uncertain information about the climate system, and the purpose of any climate modeler is to connect this information with the real world.
The branch of mathematics concerned with inference in presence of uncertain information  is nothing but statistics. 

The purpose of the present article is to review the traditional approaches to palaeoclimate modelling, and, then, to show how statistics may help us to progress in the different problems involved in the prediction and explanation of (palaeo-)climate phenomena.  The diagram on \figref{fig:concept} may be used as a road map. The different colors constitute different frameworks of palaeoclimate modelling: conceptual models, simulators (which includes general circulation models) and statistical models. They are connected to a number of concepts, which are detailed in Section \ref{sect:main}. The light gray nodes represent traditional connections between the modelling frameworks, while the dark grey ones refer to more recent concepts, introduced and commented on in Section \ref{sect:bridge}. 

\begin{FPfigure}
\tikzstyle{level 1 concept}+=[sibling angle=120]
\tikzstyle{level 2 concept}+=[sibling angle=70]
\tikzstyle{level 3 concept}+=[sibling angle=70]
\tikzstyle{level 4 concept}+=[sibling angle=70]
\tikzstyle{node}+=[mindmap]
\begin{tikzpicture}[mindmap, concept color=gray]
\path[mindmap,concept color=black,text=white]
node[concept] {Palaeoclimate Data} [clockwise from=0]
   child[concept color=green!50!black,sibling angle=30] {
     node[concept] (bs) {Statistical Models 2.3}  [clockwise from=70,sibling angle=50]
        child [concept color=green!50, text=black] { node[concept]  (ts) {Test} }
        child [concept color=green!50, text=black] { node[concept] (es) {Process Estimation}  [clockwise from=80]
             child [concept color=green!50, text=black, sibling angle=40] { node[concept] (hm) {Hierachical Modelling}  }
             child [concept color=green!50, text=black, sibling angle=40] { node[concept] (re) {Regression}  }
       }
    }
  child[concept color=blue!50!black] {
     node[concept] (c) {Conceptual Models 2.1}  [counterclockwise from=-200] 
       child {node [concept]  (sd) {Schematic description}}
       child {node [concept]   {Perceptual analogy}  
          child[concept color=blue!50!white, sibling angle=30] {node[concept] {Conveyor belt}} 
          child[concept color=blue!50!white, sibling angle=30] {node[concept] {Box model}} 
          child[concept color=blue!50!white, sibling angle=30] {node[concept] {Electronic circuit}} }
       child[concept color=blue!50!black] 
       {
        node [concept] {Mathema\-tical}
         [counterclockwise from=-120] 
         child[concept color=blue!50!white] 
            {
               node[concept] {Linear Theory 2.1.2} 
                 [counterclockwise from=-170] 
                   child { node[concept] {Forcing}  }
                   child { node[concept] {Feedback}  }
           }
         child[concept color=blue!50!white] 
            {
               node[concept] {Deterministic dynamical systems 2.1.3}    [clockwise from=-80] 
                child { node[concept] {Bifurcation Theory}   }
                child { node[concept] {Synchro\-nisation} }
            }   
         child[concept color=blue!50!white] 
            {
               node[concept] (sds) {Stochastic dynamical systems 2.1.4}    [clockwise from=-60] 
                child { node[concept] {Stochastic Resonance} }
           }   
         }
        }
     child[concept color=red!50!black] {
    node[concept] (sim) {Simulators 2.2}  [clockwise from=155] 
       child[concept color=red!70, text=black] { node[concept] (sens) {Sensitivity experiments}  [clockwise from=220]
                    child {node [concept] {parameters}} 
                    child {node [concept] {Forcing }} }
       child[concept color=red!70, text=black] { node[concept] (ev) {Evaluation}  
                    child { node [concept] {Observ. modelling} }
                    child {node [concept] (cobs) {with climate estimated from obs.}}  }
   };
\begin{pgfonlayer}{background}
   \node  [extra concept] (em) at (6, 6) {Emulator 3.2};
   \node  [extra concept] (dis) at (1, 8) {Discre\-pancy 3.1};
   \node  [extra concept, concept color=black!10, text=black] (si) at (-7, -.2) {Explanation};
  \path (sens) to[circle connection bar switch color = from (red!70) to  (black!10)] (si);
  \path (si) to[circle connection bar switch color = from (black!10) to  (blue!50!black)] (sd);
   \node  [extra concept, concept color=black!10, text=black] (rec) at (7, 9) {Recon\-struction};
  \path (dis) to[circle connection bar switch color = from (black!50) to (black!10)] (rec);
  \path (cobs) to[circle connection bar switch color = from  (red!50) to (black!10) ] (rec);
  \path (rec) to[circle connection bar switch color = from (black!10) to (green!50) ] (re);
  \node  [extra concept] (gsa) at (1.0, 3.7) {Sensitivity Analysis 3.4};
  \path (em) to[circle connection bar] (gsa);
  \path (em) to[circle connection bar] (dis);
  \path (dis) to[circle connection bar  switch color = from (black!50) to (red!50) ] (ev);
  \path (gsa) to[circle connection bar switch color = from (black!50) to  (black!10)] (si);
  \node  [extra concept] (si) at (-3.8, -.5) {System Identification 3.3};
  \path (sim) to[circle connection bar switch color = from (red!50!black) to  (black!50)] (si);
  \path (si) to[circle connection bar switch color = from (black!50) to  (blue!50)] (sds);
  \path (sim) to[circle connection bar  switch color = from (red!50!black) to  (black!50)] (em);
  \path (em) to[circle connection bar  switch color = from (black!50) to  (green!50)] (hm);
  \node  [extra concept] (pf) at (5, -6) {Dynamical system calibration 3.3};
  \path (hm) to[circle connection bar  switch color = from (green!50) to  (black!50)] (pf);
  \path (pf) to[circle connection bar  switch color = from (black!50) to  (blue!50)] (sds);

\end{pgfonlayer}
\end{tikzpicture}

\caption{Conceptual representation of the three main frameworks of palaeoclimate modelling and bridging concepts. The main frameworks are represented by the three main nodes: simulators in red,  conceptual models in blue and statistical models in green. Light grey nodes represent traditional connections between the main modelling frameworks, while the dark gray ones indicate concepts more recently introduced in the palaeoclimate modelling literature. Numbers in nodes refer to section numbers of the present article. }
\label{fig:concept}
\end{FPfigure}

\section{Traditional frameworks of palaeoclimate modelling \label{sect:main}}
\subsection{Conceptual models}
\subsubsection{Definition and construction strategy \label{sect:conconstruct}}

In general, a `conceptual model' is a drastically simplified representation of a complex process. However, even in this context, the word `model' is used diversely in the literature \footnote{cf. also \citet{Winsberg99aa} for more discussion on the meaning given to the word `model' in the context of simulation}.

\begin{figure}
\begin{center}
\includegraphics{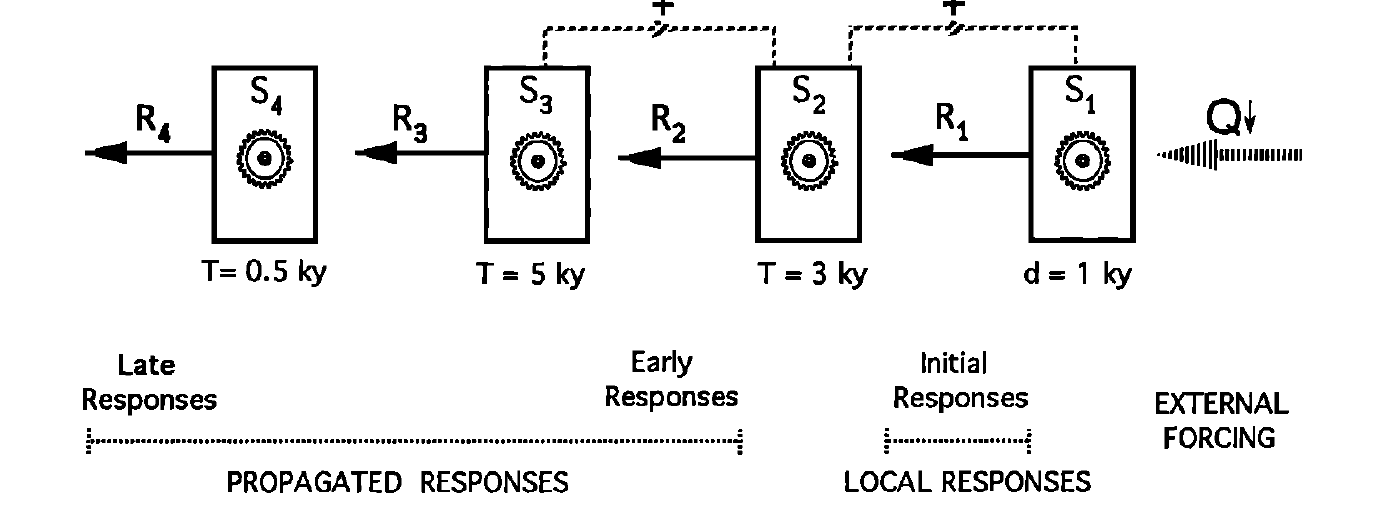}
\end{center}
\caption{The \citet{Imbrie92aa} conceptual model of the linear climate system response to the astronomical forcings associated to precession and obliquity. The model describes a sequence of initial, early and late responses noted $S_1$ to $S_4$, to the astronomical forcing ($Q$), each response lagging the previous one by 0.5 to 4~ky (1 ky=1,000 years). Reproduced with permission from American Geophysical Union. }
\label{fig:specmap}
\end{figure}

It may be a schematic description of the mechanisms involved in a complex process, often supported by a visual diagram. The `SPECMAP' model \citep{Imbrie92aa, imbrie93} is a description of the chain of responses of the different components of the Earth climate system involved in glacial-interglacial cycles \figrefp{fig:specmap}. Likewise,  the results of sensitivity experiments with general circulation models are often summarised in the form of a diagram. For example, \cite{zhao05mh} summarise a mechanistic interpretation of the effect of the ocean on the African and Indian monsoons with two diagrams outlining the roles of trade winds, evaporation and stratification.

\begin{figure}[h]
\begin{center}
\includegraphics{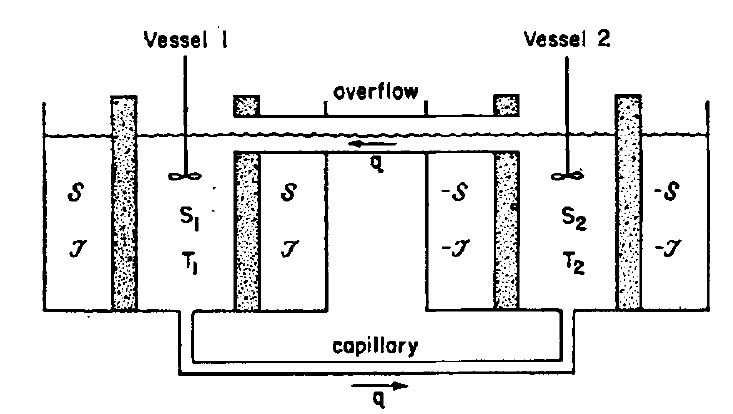}
\end{center}
\caption{Mechanical representation of the classical three-box \citet{stommel61} model, with an explicit representation of diffusive phenomena (through porous walls) and advective flows (through the capillary pipe). The Stommel model is a prototype of more complex models used nowadays (see text). Reproduced from the original publication.}
\label{fig:stommel}
\end{figure}

The word `model' may also refer to a perceptual analogy. For example, the thermohaline circulation was depicted as a `conveyor belt' \citep{Broecker91aa} \footnote{The conveyor belt logo was introduced in the November issue of the \textit{Natural History} magazine and the \citep{Broecker91aa} reference quoted here is a discussion of the relevance and limit of this analogy.}, a `leaky funnel' \citep{Mouchet08aa} and, before this, it was compared to a device that can be set up in the laboratory with two vessels and a system of stirrers, porous walls and capillary pipes materialising the effects of mixing, diffusion and advection \citep{stommel61} \figrefp{fig:stommel}. The different perceptual analogies convey different impressions about the roles of advection and diffusion in the ocean circulation physics. 
Electronic circuits are another basis for perceptual analogies. \cite{hansen84} introduced the concept of ``gain", originally used in electronics, to quantify the effects of ``feedbacks" of climate system components to a radiative perturbation. This electronic analogy was made more explicit by \citet{Schwartz11aa}, and it is further discussed later in this section \figrefp{fig:elec}. 
More abstract perceptual analogies have been suggested, in particular the relaxation model, in which it is assumed that the system is being relaxed (like a spring) to various states determined  by the forcing and the history of the system. This analogy has been much used by Paillard \citep{paillard94aa, paillard98, paillard01rge,  paillard04eps}. 

Finally,  `conceptual model' may refer to a simple system of mathematical equations. There are different strategies exist to determine these equations: 
\begin{itemize}
\item 
They may be written to translate a perceptual analogy in order to analyse and discuss its 
consequences on the system dynamics.
For example, the \cite{stommel61} model can be expressed as two equations constrained by the laws of conservation of tracers combined with a parameterisation of the response of the inter-vessel flow to the density difference between the two vessels. 
One non-trivial consequence of this model  is that it features two stable states, which have been interpreted as the on and off modes of the thermohaline circulation. Possible implications for the future of our climate are discussed in \citet{rahmstorf00thresholds}. Stommel's model is the simplest example of a wider class of mathematical models called `box models'. 
The \citet{Winton93aa} model, for example, was used to study ocean internal oscillations involved in Dansgaard-Oeschger events \citep{Schulz2002oscillators}. 
 The ``Multi-box model" (MBM) \citep{Munhoven07aa} includes 10 boxes for the ocean and was used to study climate-ocean-sediment interactions over the latest glacial-interglacial cycle. 
 The BICYCLE model, based in part on MBM, was used to interpret isotopic records \citep{Kohler05aa}. It includes 6 ocean boxes plus a number of reservoirs for the terrestrial biosphere, rocks, sediments and atmosphere.  
\item Another strategy for developing mathematical conceptual models consists in starting from fluid dynamics equations and simplify them mathematically as much as possible, using a procedure called `truncation'. The technique is well-established \citep{Saltzman62aa} and it was applied to continental ice flow dynamics to study ice ages \citep{oerlemans80,  ghil81} and Heinrich events \citep{Paillard95aa}. 
\item
Finally, the development of a conceptual model may be more heuristic, combining physical arguments, information obtained from sensitivity experiments with general circulation models and hypotheses on non-linear effects. This approach was championed by \cite{saltzman02book}.
\end{itemize}

Mathematical models may then be distinguished according to their mathematical properties, in particular, whether they are linear or non-linear, and whether they are deterministic or stochastic.
This aspect is now further developed. 

\subsubsection{Linear models \label{sect:linear}}
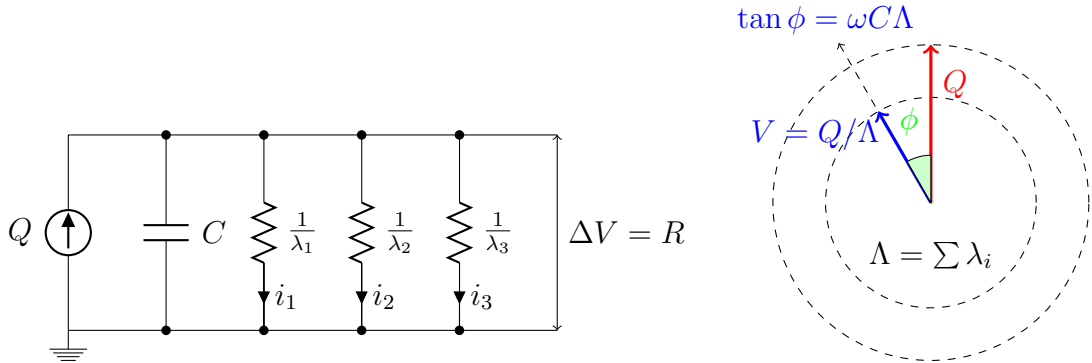
\begin{figure}
\ctikzset{bipoles/length=.9cm}
\begin{tabular}{ll}
\ctikzset{bipoles/length=1cm}
\begin{circuitikz} [american currents, scale=1.3]
 \draw  node[ground] {} (0, 0) to [I=$Q$] (0, 2)  -- (4,2);
        \draw  (1,2) to [C, l^=$C$, *-*] node[] {}  (1, 0) ; 
        \draw  (2,2) to [R, l^=$\frac{1}{\lambda_1}$ , i>=$i_1$, *-*] node[] {} (2, 0);
        \draw  (3,2) to [R, l^=$\frac{1}{\lambda_2}$ , i>=$i_2$, *-*] node[] {} (3, 0);
        \draw  (4,2) to [R, l^=$\frac{1}{\lambda_3}$ , i>=$i_3$, *-*] node[] {} (4, 0);
        \draw  (4, 2) to node  {}  (5, 2);
 \draw [<->] (5, 2) to node [right] {$\Delta V=R$} (5, 0);
 \draw (5, 0) to (0, 0);
\end{circuitikz}
& 
\begin{tikzpicture}[scale=0.7]
\draw (0, 0) [dashed] circle  (3cm);
\draw (0, 0) [dashed] circle  (2cm);
\draw[->, red, very thick] (0, 0) -- node (I) [near end, right] {$Q$} (90:3cm);
\draw[->, dashed] (0, 0) --  (120:3.5cm);
\draw[->, blue, very thick] (0, 0) -- node (V) [near end, left] {$V=Q/\Lambda$} (120:2cm);
\filldraw [fill=green!20] (0, 0) -- (0, 9mm) arc (90:120: 0.9cm)  -- cycle;
\node  [green] at (105:1.6cm)   {$\phi$}  ;
\node  [blue] at (120:4cm)   {$\tan \phi=\omega C\Lambda$}  ;
\node  at (270:1cm) {$\Lambda=\sum \lambda_i$};
\end{tikzpicture}
\end{tabular}
\caption{Linear electronic circuit as a metaphor for the climate system. The forcing is the input current $Q$. Depending on the framework, it may represent the annual mean radiative forcing (for a global warming experiment) or the net mass balance over ice sheets (for a Milankovitch forcing experiment). The capacitor represents the component of the climate system that effectively integrates the forcings (ocean heat content, or ice sheets mass) and the different resistances represent what are generally referred to as ``feedbacks" in the climate system. The values of theses resistances are consistent with the notation introduced by \cite{hansen84}: Positive resistance ($\lambda > 0$ in \cite{hansen84}) correspond to ``negative feedbacks" (they work against the forcing), negative resistances, in practice obtained with negative impedance circuits, are ``positive feedbacks". The system output (for example: temperature) is the voltage. At a given forcing frequency,  the output lags the input with a phase $\tan^{-1}(\omega C \lambda)$, where $\omega$ is the angular velocity of the forcing. }
\label{fig:elec}
\end{figure}

Consider an electronic circuit with resistors and a capacitor \figrefp{fig:elec}. Resistors dissipate current proportionally to the difference in voltage to which it is connected (the coefficient is the \textit{resistance}); and the voltage across the capacitor increases proportionally to the charge being stored on it (the coefficient is the \textit{capacity}).  The system is said \textit{linear} if the resistances and capacitance are constant.

In an analogy with the climate system, the current may viewed as the input radiation (in a global warming experiment) or the net  accumulation of snow on the ice sheets (in a Milankovitch forcing experiment). 
The capacitor accumulates charge over time. At the Milankovitch scale, ice sheets play this role, since they accumulate snow mass imbalance over several thousands of years. The growth of ice sheets causes a tension (voltage) on the system, and the role of resistors is to dissipate this tension with a discharge current, so that tension does not grow to infinity. At Milankovitch time scales, the  discharge of ice towards the ablation region and the oceans plays this role. In electronics it is also possible to design `negative resistance circuits' \citep{Linvill53aa}. Technically, this involves operational amplifiers. They inject current proportionally  to the voltage. Consequently, they amplify the effects of the forcing. In palaeoclimates, the ice albedo feedback or the boreal vegetation feedback may be modelled as negative resistances because they amplify the forcing. The combination of positive and negative resistances may be summarised by an equivalent `net resistance', as indicated on \figref{fig:elec}.

The capacitor is a crucial component of the system. It accumulates tension, like ice sheets accumulate mass. The effect of the capacitor on the system dynamics is to introduce a phase lag between the forcing (input current) and the response (output voltage), so that the forcing and the response may be unambiguously identified. The forcing-response phase lag is used by \citet{Gregory04ab} to estimate the net `feedback' factor (equivalent to the inverse of the net resistance in our diagram) in global warming experiments with general circulation model simulations. In palaeoclimate applications, the phase lag is estimated from the data to distinguish the response from the forcing \citep{Imbrie92aa,shackleton00,Lisiecki08ab}.

The assumption of linearity has three consequences.
First, given that resistances are constant, the net resistance of the system is constant, too. Consequently, it must be positive otherwise tension (response) would run away to infinity. The implication is that small perturbations to the input (for example: noisy fluctuations) are damped with time rather than being amplified.  The system remains thus predictable even in presence of small fluctuations. 
Second, the output voltage contains exactly the same frequencies as the input current, and every frequency can be analysed separately from the others. The  technique introduced in the SPECMAP project \citep{Imbrie92aa} is compatible with this linear framework. It consists in filtering climatic signals to separate them  into components related to precession (19-24 ky band, 1~ky=1,000 years), obliquity (40-ky band) and eccentricity (100-ky band), and then inspect leads and lags between the astronomical forcing and the different climatic components, such as CO$_2$, ocean temperatures, dust, etc. Visually, the picture proposed by SPECMAP is a chain linking the forcing to different components, of which the response has an increasing phase lag with the forcing as one goes down the response chain \figrefp{fig:specmap}. This analysis approach is still used nowadays \citep[e.g.][]{Lisiecki08ab}.

Third, the linear formalism cannot adequately account for the presence of saw-tooth-shaped 100-ky ice age cycles. At least three answers to this problem may be found in the literature:
\begin{enumerate}
\item The first solution consists in postulating the existence of some unidentified 100-ky cycle that is effectively treated as a forcing in the linear framework, even if its origin is internal to the climate system. This is the solution proposed in the SPECMAP model \citep{imbrie93}.
\item The second approach assumes that the effect of the forcing is non linear. For example, in \citet{imbrie80}, the response to insolation anomaly is greater if this anomaly is positive than if it is negative. 
The consequence of this asymmetry is that the input signal is at least partly rectified. In particular, the precession forcing, which is the product of eccentricity and the sine of the longitude of the perihelion ($e \times \sin\varpi$) gives rise to a pure eccentricity signal in the output. The eccentricity signal contains the frequencies of 412, 94, 123 and 100~ky \citep{berger78}, and therefore these frequencies are also found in the output signal. 
\citet{Ruddiman06aa} proposed a conceptual model which, although not formulated mathematically, fits this category. 
The problem with this second approach is that the internal dynamics of the system remain linear. Only the effect of the forcing is being transformed non-linearly. Consequently, as we discussed above, there is no possibility of a runaway feedback phenomenon. This causes a problem to explain the deglaciation which occurred 400,000 years ago because insolation variations were weak at that time. If the system is stable, it cannot react dramatically to a small forcing. The problem was fully appreciated by \citet{imbrie80}. 

\item The third solution is to explain the emergence of an `internal 100-ky cycle' as a result of  non-linear interactions with the climate system. This is the subject of the next section.  
\end{enumerate}

\subsubsection{Non-linear, deterministic models}
\begin{figure}
\begin{center}
(a) Saltzman and Maash (1991) model \newline
\frame{
\begin{tikzpicture}[->,>=stealth, auto, semithick]
\tikzstyle{every node}=[draw]
\node (F)  at (-3,1)    {Forcing} ;
\tikzstyle{every node}=[circle, draw, minimum size=3em]
\node (I)  at (0,0)    {\parbox{3em}{Ice Mass} };
\node (O)  at (2,2)    {\parbox{3em}{Ocean Deep Temp.} };
\node (C)  at (4,0)    {\parbox{3em}{CO$_2$} };
\tikzstyle{every node}=[draw=none, minimum size=0cm]
\path  (F) edge [bend left] (I) ;
\path  (I) edge [loop below] node (p) [below] {-} (I) ;
\path  (I) edge [bend left] node (p) [above] {-} (O) ;
\path  (O) edge [bend left] node (p) [above] {-} (I) ;
\path  (C) edge [loop right] node (p) [right] {+/-} (C) ;
\path  (O) edge [loop above] node (p) [above] {-} (O) ;
\path  (C) edge [bend left] node (p) [above] {-} (I) ;
\path  (O) edge [bend left] node (p) [above] {+/-} (C) ;
\end{tikzpicture}
}
\begin{tabular}{ll}
\includegraphics[scale=0.5]{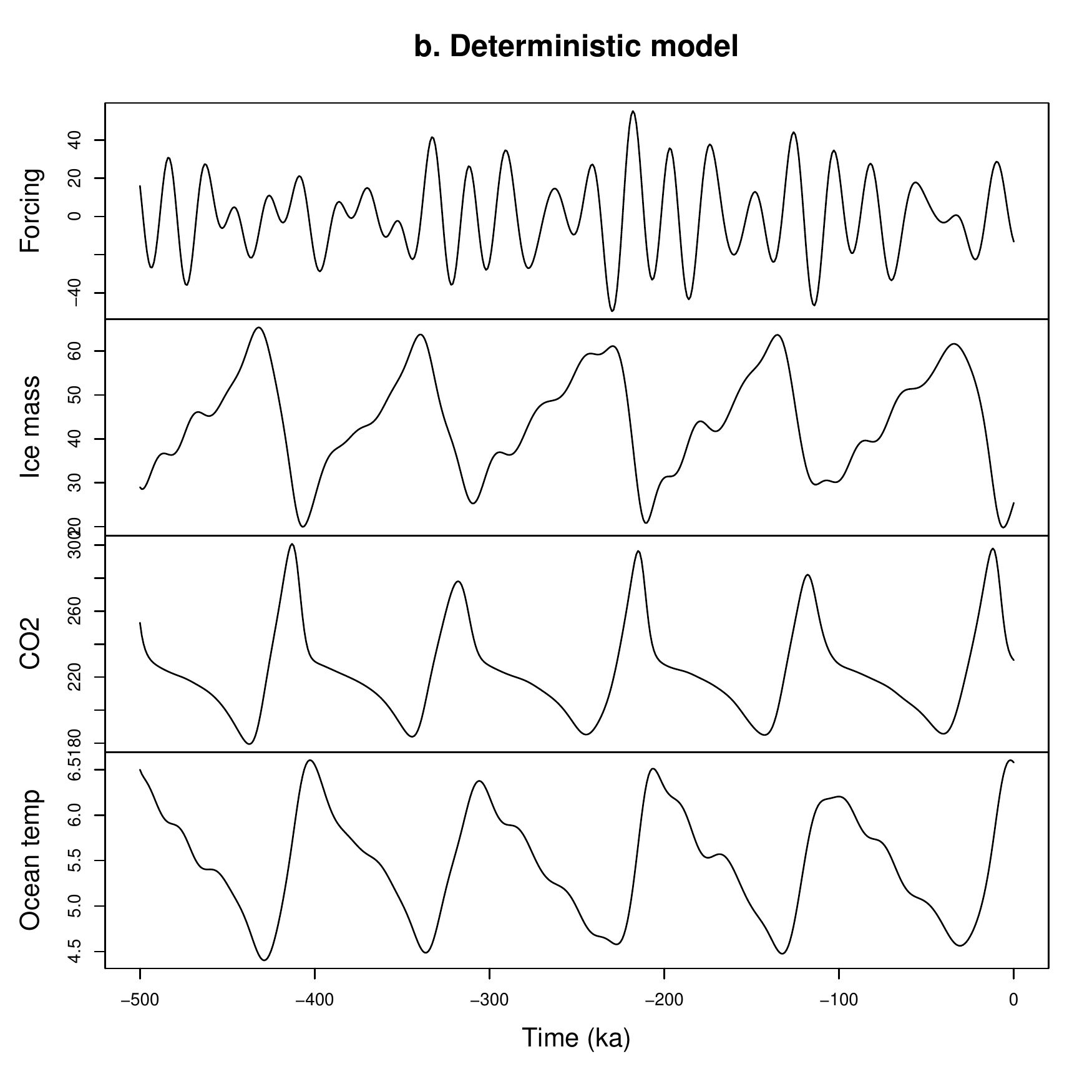} & 
\includegraphics[scale=0.5]{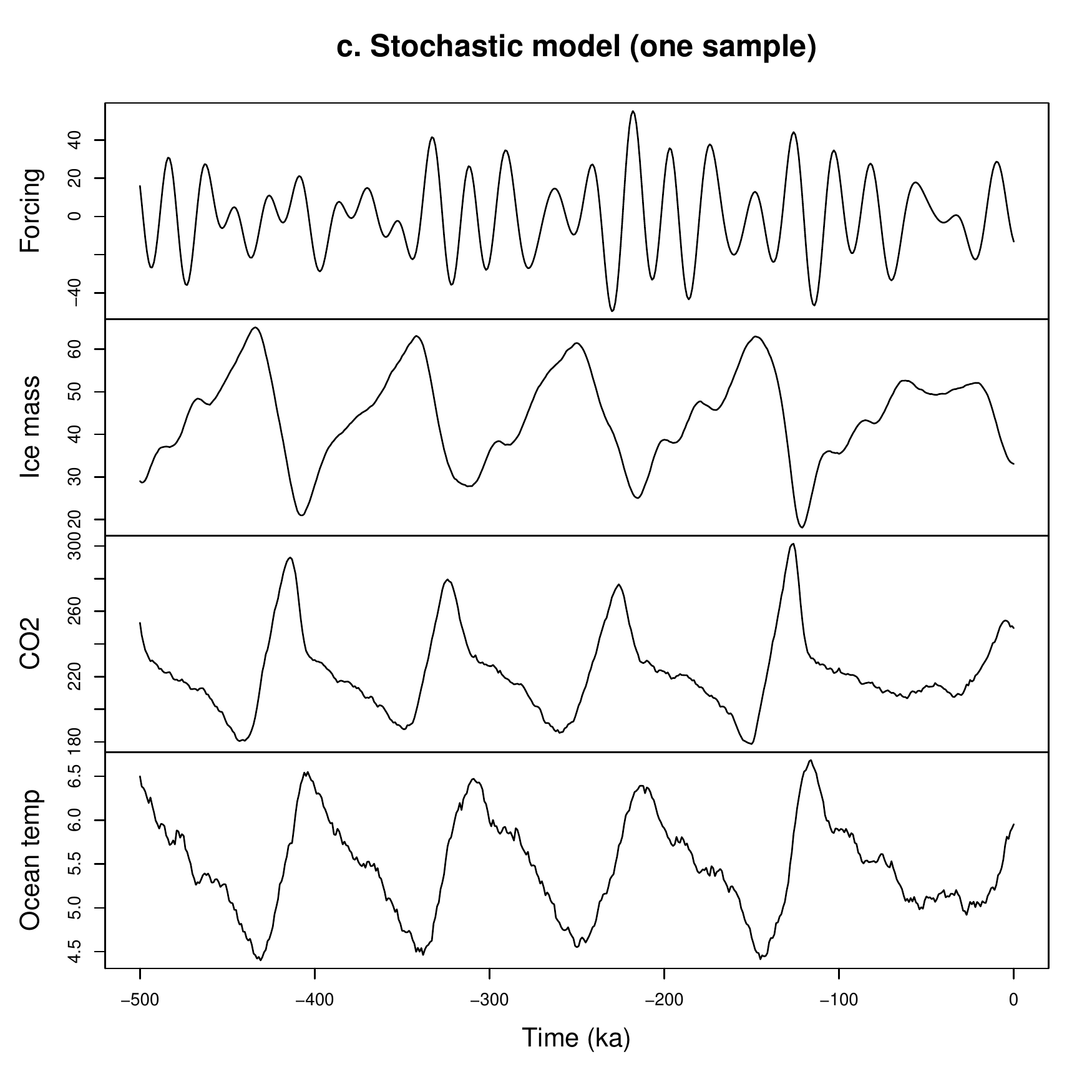} \\
\end{tabular}
\end{center}
\caption{(a) Graphical network representation of the \citet{saltzman90sm} model of ice ages. Arrows indicate \textit{causal} relationships. The astronomical forcing causes variations in ice volume, which then are involved in a network of interactions with deep ocean temperature and CO$_2$. Contrary to the linear framework, certain feedbacks may vary and change sign depending on the system state, which induces in the \citet{saltzman90sm} model a phenomenon of self-sustained oscillations. Ice ages are, in this model, self-sustained oscillations synchronised on the astronomical forcing. The result of a simulation reproducing Figure 5 of the original publication is shown in panel (b). Additive stochastic effects may randomly alter the timing and amplitude of ice ages. One example of a stochastic realisation is shown on panel (c).}
\label{fig:saltzman}
\end{figure}

A non-linear electronic system is one in which the effective resistance of the components vary with the tension on the system. 
This framework allows for a variety of new phenomena. 
In ``some basic experiments with a vertically-integrated ice sheet model", \cite{Oerlemans81aa} noted one,   two or even three stable solutions may co-exist, depending on the forcing. This configuration, typical of non-linear system dynamics, may cause transient behaviours that will look very different to the  forcing functions. In particular, with some  modest adjustments to the model equations and parameters (as in \cite{Pollard83aa}),  such a model may exhibit self-sustained 
oscillations, even if the forcing is constant. In non-linear dynamics theory, such a system is called an \textit{oscillator}.  In the \citet{Pollard83aa} model, the oscillation arises from interactions between the lithosphere and the ice-sheet climate system, but only when model parameters lie in a fairly narrow  range, that is suitable to represent the West Antarctic ice sheet but not the northern ice sheets. Consequently, investigators have been looking for other sources of oscillations in the climate system to explain ice ages.  \citet{saltzman90sm}, for example, proposed a system in which CO$_2$ reacts non-linearly to changes in ocean temperature, and ocean temperature is controlled by CO$_2$ and continental ice volume. Contrary to the SPECMAP model, which is graphically represented as a linear chain of responses \figrefp{fig:specmap}, the Saltzman-Maasch model is best graphically represented as a network with cyclic interactions \figrefp{fig:saltzman}. Some of these interactions act as stabilising factors (negative feedbacks, noted `$-$'), but others may be stabilising or destabilising depending on the system state. 
This latter characteristic  allows, in the Saltzman-Maasch model, ice ages to occur even in absence of astronomical forcing. 
Consequently, in a non-linear system, the phase relationship between two components (for example, astronomical forcing and ice volume) is no longer a reliable indicator of the nature of the forcing (see also \cite{Ganopolski09aa}). Compared to linear systems, the analysis focuses less on the analysis of phase lead and lag, and is more concerned with identification of  stability, bifurcation, of synchronisation, and predictability, which are now briefly reviewed.

\begin{description}
\item[Stability:] The analysis of a non-linear dynamical system consists in identifying its stable and unstable points, that is, the states to which the system may be attracted (negative feedbacks dominate around this point) and the states from which it will be repelled (positive feedbacks dominate around this point). 
For example, \cite{Brovkin98aa, brovkin03cc} estimated that non-linear interaction between vegetation and the atmosphere may explain the co-existence of several stable states in the Sahel but not at the northern high latitudes; this work was inspired in part from the pioneering contribution by \cite{Ghil76aa}, interested in the more general consequences of albedo feedbacks on the system stability.
There may also be no stable point, which leaves us with several possibilities \footnote{see, among others, \cite{Ghil87aa,  saltzman02book} and \citet{Crucifix12aa} for introductory texts on this in the palaeoclimate context}:  (i)  the system exhibits a stable self-sustained periodic oscillation; (ii) its behaviour is quasi-periodic (combination of several periods); (iii) it is aperiodic, with a broad power spectrum and complex phase-diagram figures (for example: the behaviour of ENSO modelled in \cite{tziperman94}).

\item[Bifurcation:] 
A bifurcation is defined as a change in system behaviour obtained when one of the system parameters crosses a threshold,  called the bifurcation point.  \citet{saltzman90sm}, for example, proposed models where the Middle Pleistocene Revolution is interpreted as a bifurcation. The parameter controlling this bifurcation is the background CO$_2$ level, which is driven by tectonics. This bifurcation corresponds to a switch from a linear response mode to astronomical forcing (before the Middle Pleistocene Revolution) to a regime of non-linear synchronisation (after the Middle Pleistocene Revolution). This model is compatible with the findings of \citet{lisiecki07trends} who show, based on time series analysis, a transition from linear to non-linear regime before the Middle  Pleistocene Revolution. At a different timescale, the transition between an `interglacial'  and a `glacial' phase may also be interpreted in terms of bifurcation theory \citep{Ditlevsen09aa, Livina11aa} \footnote{The difference between the \cite{Ditlevsen09aa} and \cite{saltzman90sm} frameworks is quite fundamental but its discussion is beyond the scope of the present review.}. Recent research activity has focused on the search for `early warning precursors' of a bifurcation \cite{Dakos08aa,  Scheffer09ab, Ditlevsen10aa}. 
\item[Synchronisation:]
Synchronisation is defined as the phenomenon by which the natural oscillation period of  a system adjusts itself on the period of an external factor \citep{Pikovski01aa}.
Conceptual models such as those by \citet{saltzman90sm}, \citet{paillard04eps} and \citet{tziperman06pacing} exhibit a phenomenon of synchronisation to the astronomical forcing 
\footnote{
The Southern Ocean foraminifera Fourier analysis by \cite{hays76} is often referred to as a pioneering demonstration of the influence of the astronomical forcing on climate. Retrospectively, it is tempting to interpret the title of that paper: ``Variations in the Earth's orbit, pacemaker of ice ages'' as a visionary reference to the concept of synchronisation. \cite{hays76} indeed perfectly realised that occurrence of 100-ka ice ages does not fit a linear response framework of precession and obliquity. Though,  as correctly observed by one reviewer of the present paper, their interpretation is closer in nature to the \cite{imbrie80} model than of the self-sustained oscillation model.}
The concept of synchronisation is also useful to explain phenomena associated with Dansgaard-Oeschger and Heinrich events \citep{Schulz2002oscillators}. A review is available in \cite{Crucifix12aa}.
\item[Predictability:] Given that positive feedbacks may dominate at least at certain times in a non-linear systems, there is the possibility small perturbations  be amplified so much that the exact history of the system from given initial conditions is in practice unpredictable.  Some conceptual models of ice ages have this property. More specifically, climate perturbations occurring at strategic times may be amplified, and inflect the course of climate, by hastening or delaying significantly a glacial inception or a deglaciation \citep{De-Saedeleer12aa}.  
\end{description}

In summary, non-linear deterministic models widen the scope of plausible ice age theories and address questions at a different level than linear systems. The idea of a `chronological chain of response mechanisms' is ambiguous in such systems because self-sustained loops are possible. Non-linear dynamical system theory provides a suitable vocabulary to study the effects of a network of interactions between different components of the climate system and the role of astronomical forcing.

\subsubsection{Stochastic models}
A stochastic process is a form of \textit{noise}, that is, the succession of random numbers with well-defined properties such as auto-correlation and amplitude. There are two possible reasons for introducing a stochastic process in a model. One is to account for the physical phenomena that occur at smaller spatial or temporal scales than those that are  explicitly resolved by the model.
For example, \citet{Hasselmann76} introduced a stochastic process to account for `weather' in a `climate' model. The properties of the stochastic process may then be determined by reasoning on the physics of the unresolved process \citep{Penland07aa}. For representing weather, it was proposed to deduce noise properties from quasi-geostrophic theory \citep{Majda09aa}. Stochastic parameterisations may also be introduced to account for uncertainties and model errors, in order to more realistically assess the horizon of predictability of the model in the presence of such errors.  Technically, accounting for stochastic effects transforms the dynamical system into a `stochastic dynamical system', of which the numerical resolution is both theoretically and computationally more involved than that of a deterministic system \citep{Kloeden99aa}. Stochastic processes also induce a number of interesting mathematical phenomena, which in turn provide us with a series of new concepts that may be adequate to interpret the history of climate. One such concept is the `stochastic resonance', in which the stochastic process is combined with a periodic forcing to induce periodic oscillations in the system. Ice ages have been interpreted in the past as a phenomenon of stochastic resonance \citep{BENZI82aa}. The concept of stochastic resonance itself has evolved and models showing different forms of stochastic resonance have been proposed to understand abrupt climate events \citep{Ganopolski2002prl, Timmermann03ab, Braun07aa}. 
\subsection{Climate simulators}
Simulators are complex numerical systems designed to account explicitly for a large number of 
different processes involved in the dynamics of the climate system. 
A particularity of simulators is that the number of variables is several order of magnitudes greater than the number of system parameters \cite{claussen02emic}.  This criteria is met by general circulation models as those used, for example, in the Palaeoclimate Modelling Intercomparison Project \citep{braconnot07pmip1}, but the definition applies also to most `Earth System Models of Intermediate Complexity (EMICS)' \citep{claussen02emic}, such as, among others, FAMOUS \citep{Smith12aa}, LOVECLIM \citep{Goosse10aa}, CLIMBER-2 \citep{petoukhov00, Calov2005Transient-simul}, CLIMBER-3$\alpha$ \citep{Montoya05aa}, BERN3D \citep{Ritz10ab} and GENIE-1 \citep{Ridgwell07aa}.

The word `simulator' is uncommon in the climate literature. It is adopted here on purpose to distinguish it from statistical and conceptual models. Compared to the conceptual models, the purpose of the simulator is more general, in the sense that it is developed with many possible applications in mind (a same climate simulator may be used to study El-Ni\~no and sea-ice variability; as another example, the GENIE-1 model \citep{Ridgwell07aa} contains the definition of 49 dissolved tracers,  which can be enabled or disabled depending on the modeller's purpose). At the same time, experimental designs are quite specific: among others, the shapes and orography of continents, vegetation types, soil properties and ocean bathymetry have to be specified on a given grid and adapted for a particular climate era, for example the pre-industrial era or the Last Glacial Maximum. 

The interest of simulators is to generate a self-consistent picture of regional and planetary-scale phenomena that is compatible with physical and biogeochemical principles implemented at the scale of a grid-cell. 
A large fraction of computing resources is used to solve fluid dynamics equations that determine the movements of oceans, atmosphere, and possibly sea-ice and continental ice. The choice of spatial resolution conditions the spatio-temporal spectrum that can be studied with the simulator, ranging from weather forecast (hi-resolution models) to multi-millennial phenomena (EMICS). Physical and biogeochemical processes are then introduced consistently. For example, a module of ocean sediment diagenesis  finds its place in a simulator designed to study multi-millennial climate evolution \citep{Brovkin12aa}, but not in a weather forecast simulator. 

The simulator also involves a number of `sub-grid parameterisations'. 
These are models representing the effects of physical or biogeochemical processes that are not explicitly resolved by the simulator. The development of these parameterisations obeys the same rules as that of the conceptual models outlined above, and they may be diagnostic (assume an instantaneous equilibrium response), prognostic (the parameterisation is a dynamical system, for example, the prognostic cloud scheme of the UK Met Office model \citep{Wilson08aa}) and they may also include stochastic components \citep{Eckermann11aa}.
Sub-grid cell parameterisations contain parameters that cannot directly be measured in the laboratory or in the field \citep{Palmer05aa}, but  plausible ranges can be estimated from physical considerations. 
Consequently, at least some of these parameters are subsequently `tuned' so that the simulator replicates satisfactorily planetary scale phenomena. For example, they may be adjusted to get a satisfactory thermohaline circulation. 
In this sense, the simulator may be viewed as a device that constraints relationships between information formulated at different time and spatial levels, namely the grid cell and the planetary scale. In addition to sub-grid parameterisations, simulators often involve a number of reasonably ad-hoc assumptions, for example about the routing of icebergs in the Atlantic ocean and the amount of freshwater subsequently delivered to the ocean surface \citep[][p. 150]{gordon00}. 

There are different purposes to the use of simulators in palaeoclimatology. 
\begin{enumerate}
\item One is to evaluate the simulator. 
The purpose is to show that the simulator reproduces palaeoclimate observations satisfactorily (or, at least, more satisfactorily than another one), and take this as an element of confidence into future climate predictions with this simulator, in the context of the anthropogenic climate change problematic \citep{braconnot07pmip1,Braconnot12aa}.  
To this end, experiments are designed carefully to be as realistic as possible. 
\item
Another purpose is to construct an explanatory framework to a climate phenomenon. For example, explaining the desertification of the Sahara 6,000 years ago \citep{Kutzbach97, claussen99}, the strong East-Asian monsoon signal during marine isotopic stage 13 \citep{yin08}, the timing of the Holocene climate optimum \citep{Renssen09aa} or the effect of different oceanic factors to variations in CO$_2$ \citep{Archer00aa, Chikamoto12aa}. Given the nature of simulators, one single simulation cannot count as a fully satisfactory explanation of a phenomenon. The interest of simulators lies in the possibility of designing series of ``intervention" experiments (change the forcing, or take the control of certain model components, such as vegetation) in order to investigate causal relationships. Even though simulators do not replicate reality exactly, these experiments 
help us to quantify complex causal effects, and possibly summarise them as a visual flow diagram (cf. Section \ref{sect:conconstruct}). 
\item The third purpose is to generate information that is not immediately accessible from palaeoclimate archives. Simulator experiments may be used to provide climate reconstructions \citep{Paul05aa} or constrain unknown quantities, such as the duration of hydrological perturbations associated with Heinrich Events \citep{Roche04aa}.
\end{enumerate}
The simulator is an imperfect representation of reality and therefore cannot replicate observations exactly. It is never `true' \citep{Oreskes04021994}. Consequently, modellers have been looking at ways of expressing the distance between the simulation and the reality, either qualitatively or quantitatively. 
The classical procedure is illustrated in \figref{fig:pmipway}. The log of data produced by the simulation is post-processed and summarised, for example in the form of seasonal averages and maps. In parallel, palaeoclimate observations are compiled and expressed in a model-friendly (or modeller-friendly!) database, that is, they are expressed in terms of climate variables, using statistical or mechanical relationships, and aggregated on a grid (an example is the MARGO dataset \citep{margo09aa}). Finally, the distance between observations and simulation may be discussed either qualitatively \citep{Otto-Bliesner09aa}, or on the basis of formal metrics.  For example, a distance metric based on fuzzy logic was proposed in the late 1990's \citep{Guiot99aa}. This  particular metric was designed to avoid  penalizing excessively a simulator that would reproduce climate change patterns correctly, but with shifts in location compared to reality. 
The model-data comparison procedure is an integral part of the modelling process. More specifically, which quantities are being looked at (seasonal averages, regional averages) and which metrics are being chosen, reflect the judgements of the modeller about which information generated by the simulator is potentially valuable for inference (cf. \citep{Guillemot10ab} for the science historian prospective of  on this matter).

From there research efforts have taken two complementary (and non-exclusive) directions:
\begin{enumerate}
\item Develop process-based models for palaeoclimate observations, so that the simulator generates information of which the nature is similar to the one which is observed. The process is illustrated on \figref{fig:proxymodel}. Rather than comparing simulated climate with reconstructed climate, one compares simulated isotopes with observed isotopes \citep{Hoffmann98aa,Marchal2000path,legrande06isotopes,  Roche06aa,Muller06aa} or biomes with pollen scores \citep{Harrison03lgm,Brewer07aa}. As for climate models, process-based models for palaeoclimate observations are available with various levels of sophistication, and they can be intimately embedded in the climate simulator (necessary for isotopes simulation) or included as a separate post-processing module (e.g.: the BIOME model, \cite{Kaplan03aa}). Compared to the classical procedure of climate-climate comparison,  observation modelling mitigates the loss of information occurring when observations are converted into climatic variables. The modeller also has a better control on the assumptions linking climatic variables to palaeo-observations. 
\item Develop sampling strategies, founded on ensembles of experiments. For example one may sample different parameter combinations of a simulator in order to explore the effect of uncertainties on the value of these parameters. The parameter combinations giving simulations comparing most favourably to palaeoclimate data may provide more trustworthy predictions \citep{annan05sola, schneidervd06lgm}. Simulations obtained with these best parameters may also be viewed as a form of climate reconstruction \citep{Paul05aa}. Sampling forcing scenarios may also provide constraints on the latter \citep{Roche06aa}. With the ensemble strategy, the interest of modellers has moved from the notion of evaluating one experiment to evaluating \textit{the experiment ensemble}, that is, asking whether climate summaries (e.g.: zonal averaged temperature) generated  by the ensemble is likely to encompass the true value \citep{Hargreaves11aa}. Section \ref{sect:bridge} provides more discussion on how an ensemble-based strategy fits a statistical approach.
\end{enumerate}

\begin{figure}
\begin{center}
{\large\bf  model-data comparison }\\
\bigskip
\begin{tikzpicture}
\tikzstyle{every node}=[fill=blue!20, rounded corners, draw=black!50, thick]
\node (param)     at (-2,4)  {\parbox{8em}{(uncertain) \\ parameters}};
\node (bound)     at (2,4)  {\parbox{8em}{Forcing and boundary conditions}};
\node (gcm)       at (0,2)  {GCM simulation};
\node (gcmsum)   at (0,0)  {summary (e.g. maps)};
\draw[->] (param) -> (gcm);
\draw[->] (bound) -> (gcm);
\draw[->, color=red] (gcm) -> (gcmsum);

\tikzstyle{every node}=[fill=green!20, rounded corners, draw=black!50, thick]
\node (paleoobs)                at (7.5,4)  {paleoobservation};
\node (aggregation)             at (7.5,2)  {compilation};
\node (climrec)     at (7.5,0)  {\parbox{8em}{Climate \\ reconstruction}};

\draw[->] (paleoobs) -> (aggregation);
\draw[->] (aggregation) -> (climrec);

\tikzstyle{every node}=[fill=white, draw=black, rounded corners]
\node [text=red, draw=red] (metric) at (4.2, -1.5) {metric};
\draw[->, color=red] (climrec) -> (metric);
\draw[->, color=red] (gcmsum) -> (metric);

\draw [dashed] (-3.9, -0.7) -- (3.9, -0.7) -- (3.9, 5) -- (-3.9, 5) -- (-3.9, -.7) ;
\draw [dashed] (5.3, -0.7) -- (10.3, -0.7) -- (10.3, 5) -- (5.3, 5) -- (5.3, -.7) ;
\end{tikzpicture}
\caption{Classically, simulations and data-based palaeoclimate reconstructions are developed separately and then compared, either based on visual inspection, or using quantitative functions (metrics) measuring the distance between the simulation and the palaeoclimate reconstruction. }
\label{fig:pmipway} 
\end{center}
\end{figure}
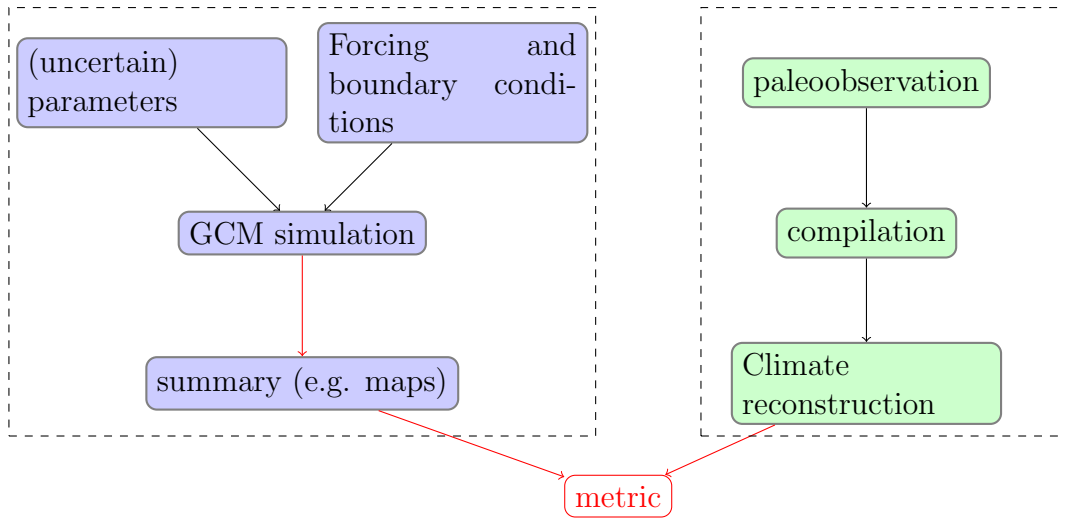
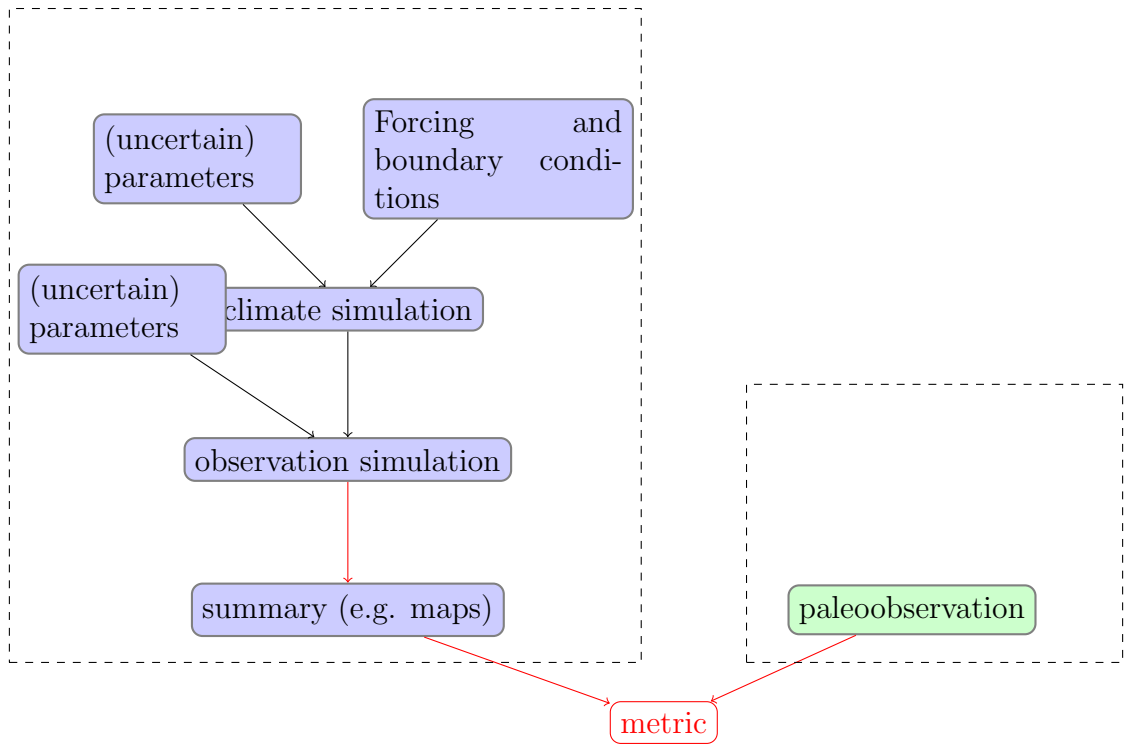
\begin{figure}
\begin{center}
{\large\bf  simulation of observations}\\
\bigskip
\begin{tikzpicture}
\tikzstyle{every node}=[fill=blue!20, rounded corners, draw=black!50, thick]
\node (param)     at (-2,6)  {\parbox{6em}{(uncertain) \\ parameters}};
\node (bound)     at (2,6)  {\parbox{8em}{Forcing and boundary conditions}};
\node (gcm)       at (0,4)  {climate simulation};
\node (paramp)     at (-3,4)  {\parbox{6em}{(uncertain) \\ parameters}};
\node (proxysim)   at (0,2)  {observation simulation};
\node (proxysum)   at (0, 0)  {summary (e.g. maps)};
\draw[->] (param) -> (gcm);
\draw[->] (bound) -> (gcm);
\draw[->] (paramp) -> (proxysim);
\draw[->] (gcm) -> (proxysim);
\draw[->, color=red] (proxysim) -> (proxysum);

\tikzstyle{every node}=[fill=green!20, rounded corners, draw=black!50, thick]
\node (paleoobs)                at (7.5,0)  {paleoobservation};
\tikzstyle{every node}=[fill=white, draw=black, rounded corners]
\node [text=red, draw=red] (metric) at (4.2, -1.5) {metric};
\draw[->, color=red] (paleoobs) -> (metric);
\draw[->, color=red] (proxysum) -> (metric);

\draw [dashed] (-4.5, -0.7) -- (3.9, -0.7) -- (3.9, 8) -- (-4.5, 8) -- (-4.5, -.7) ;
\draw [dashed] (5.3, -0.7) -- (10.3, -0.7) -- (10.3, 3) -- (5.3, 3) -- (5.3, -.7) ;

\end{tikzpicture}
%

\end{center}
\caption{As an adaptation of the methodology shown in \figref{fig:pmipway}, it is also possible to simulate palaeo-observations using appropriate models, for example, vegetation or isotope models, so that the simulation-observation comparison is done at the level of the recorded information, rather than at the level of climatic variables.}
\label{fig:proxymodel}
\end{figure}

\subsection{Statistical modelling}
A statistical model is a mathematical construction that accommodates both systematic relationships between different variables and random effects \citep{Davison03aa}, and random effects are modelled through the notion of probability distributions. 

In classical statistics, one generally distinguishes two objectives: testing a hypothesis, and estimating a quantity. In both cases, the inference procedure relies on the definition of quantities called `statistics', and on the specification of a process that could plausibly have generated the observations.

For example, \cite{huybers05obliquity} ask whether deglaciations occur systematically for a same phase of obliquity. To this end they consider a statistic called the ``R" statistic, and they study the distribution of this quantity that would have been obtained if the process that generated the benthic record had been replicated a large number of times, accounting for uncertainties on the chronology. They then observe that this hypothesis \textit{cannot} be rejected but another hypothesis, that obliquity phases at terminations are totally random, \textit{can} be rejected. 
In another application, \cite{Braun11aa} use a two-parameter process to test a possible solar influence on triggering abrupt events, and they use the `mean waiting time' as the test statistic. As a last example, \cite{Haam10aa} developed a statistical algorithm to test the correlation between two time series with chronological uncertainties. 

So far,  most problems of estimation in Quaternary palaeoclimatology were posed as a problem of regression, in which it is assumed that an observation is the sum of a systematic process and a random component. Namely, observations may be modelled as the sum of a climate-controlled component and a random observation error. Depending on the nature of the relationship, the regression problem may be solved with more or less complex techniques, including linear regression, neural networks and ``best analog" approaches  \cite[e.g.][]{kucera05margo}. The problem of interpolating sparse observations on a grid may also be posed as a regression problem, and it involves then a model of the spatial dependency of observations. Discussion and application on sea-surface temperature estimates of the Last Glacial Maximum are available in \citet{Schafer-Neth05aa}. Finally, specific time-process models have been tailored for palaeoclimate time-series analysis. 
In particular, the `RAMPFIT' model \citep{Manfred00aa} assumes a linear trend between two plateaus. It was used to estimate the timing of the Northern Hemisphere glaciation based on foraminifera data \citep{Mudelsee05aa}.

More recently, palaeoclimate scientists have become interested in modelling probability distributions of uncertain variables, and the Bayesian approach provides a framework to this end.
The principle is the following. Observations are modelled as the output of a process expressed mathematically. The process generally involves a number of parameters. 
In the Bayesian formalism, knowledge on the natural process is being incorporated by different channels. The first one is the structure of the equations that represent the process. Examples are given later in this review, in which sophisticated vegetation or climate models are used to this end; though the modeller may chose a very general model. For example, auto-regressive processes are a very broad class of models that express the fact that the system has a memory. Physical knowledge also concerns the range of values that parameters may plausibly take. This is the second channel. This information is incorporated in the Bayesian statistical model under the form of \textit{prior} probability distributions of parameters, which reflect the judgements of the modeller about the probability that the parameters will take this or that value.
Finally, observations are accounted for
in a mathematical procedure that consists in  updating the prior distributions on parameters by application of the Bayes' law of conditional probabilities. This operation is achieved by resorting to a function called the likelihood. This is a function of the parameters, which expresses the degree of consistency between the model output obtained with these parameters, and observations at hand \footnote{Technical definitions of the likelihood and explanation of Bayesian calculus are available in classical textbooks, such as \cite{Davison03aa}, but the broader discussion by \cite[][chap. 3]{jaynes03aa} is also quite enlightening.}. The resulting probability distribution of the parameters, which combines prior and likelihood, is called the \textit{posterior} probability distribution. 

The Bayesian formalism is also equipped with a concept to quantify the degree of self-consistency between these different informations. The relevant quantity is called the marginal likelihood: Among different models,  the one with the highest marginal likelihood is generally to be preferred. Though, this is by no means automatic. Assessing a model remains a broad and partly subjective procedure that involves testing it against independent data, and discussing its structure based on physical considerations. 

Bayesian models were proposed to estimate  distributions of sediment age based on uncertain radiocarbon dates \citep{Blaauw05aa,Haslett08aa},  infer the posterior distribution of Holocene temperatures based on chironomid taxa abundances \citep{Korhola02aa}, or estimate ice core chronologies \citep{Lemieux-Dudon10aa}.

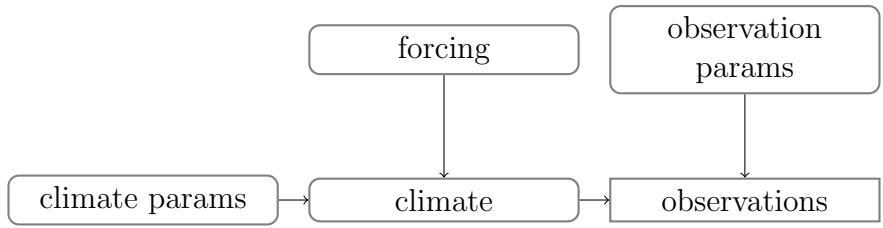
\begin{figure}
\begin{center}
{\large\bf Hierarchical statistical modelling }\\
\bigskip
\begin{tikzpicture}
\tikzstyle{every node}=[text centered, rounded corners, text width=8em, draw=black!50, thick]
\node (forcing)     at (0,2)  {forcing };
\node (scov)     at (-4,0)  {climate params };
\node (climate)     at (0,0)  {climate};
\node (proxy)     at (4, 2) {observation params};
\node (obs) [sharp corners]      at (4,0)  {observations};
\draw[->] (forcing) -> (climate); 
\draw[->] (scov) -> (climate); 
\draw[->] (climate) -> (obs);
\draw[->] (proxy) -> (obs);
\end{tikzpicture}
\end{center}
\caption{The hierarchical model is a statistical model, which can be represented graphically using a graphical network as shown here. The purpose of the network is to represent what each variable depends statistically on. In this example, the state of climate depends on parameters (describing climate processes) and on the forcing. What is being actually observed (observations) depends on climate and on a number of parameters that control the way climate information is being stored in climate's natural archive.
Statistical inference consists in updating our knowledge of climate and parameters, given observations and knowledge of the forcing. }
\label{fig:statmodel}
\end{figure}

The Bayesian approach also provides the flexibility needed to construct `hierarchical models' \figrefp{fig:statmodel}. The hierarchical model chains different sub-models to form a consistent forward description of the process that leads to the observation. For example, it is possible to articulate a  process to model response of climate to a forcing, with one model that expresses the response of a climate record to climate (for example: $\delta^{18}$O in a marine sediment). The statistical resolution of the problem consists in finding  the joined probability distribution of all variables involved in the model, based on prior evidence and observations. 

Hierarchical modelling was recently introduced in  palaeoclimate applications. 
\cite{Haslett06aa} use a hierarchical model to reconstruct climate from pollen samples on the site of Glendalough in Ireland. \cite{Li10aa} apply a hierarchical model to estimate the evolution of annual mean, northern hemisphere temperature during the last millennium based on three observation time series and three estimates of the climate forcing (volcanoes, solar and greenhouse gases), and \cite{Tingley10aa} combined model for the geographic correlation and correlation of climate states in a model, which they then chained with a model for the observations. They called this model `BARCAST', for ``Bayesian Algorithm for Reconstructing Climate Anomalies in Space and Time", and they compare it with different other approaches, Bayesian or not, in \cite{Tingley12aa}.

\section{Combining statistical with process-based modelling \label{sect:bridge}}
\subsection{Principle}
The purpose of process-based modelling is to articulate physical and biogeochemical principles, be they deterministic or stochastic. Statistical modelling, on the other hand, is concerned with inference about uncertain variables. 
Here we show how these two approaches may be combined. To illustrate the principle, consider the work of  \citet{Guiot2000Inverse-vegetat}. In this study, the BIOME3 process-based vegetation model \citep{Haxeltine96aa} is used to generate  prior distributions of the state vegetation in modern and past climates. The prior is then updated based on observed pollen fossil samples. The approach is thus statistical in principle, but physical or biogeochimical principles are used to restrict the range of scenarios that may be admitted as plausible. Likewise, \citet{annan05sola} considered a range of parameters of the climate simulator MIROC2 to generate a plausible probability distribution of the climate of the Last Glacial Maximum. They obtained a posterior distribution of the parameters of that model, which they used to construct a posterior distribution of climate sensitivity to anthropogenic perturbations.

\begin{figure}
\begin{center}
{\large\bf Simulator-assisted probabilistic-modelling }\\
\bigskip
\begin{tikzpicture}
\tikzstyle{every node}=[fill=blue!20, rounded corners, draw=black!50, thick]
\tikzstyle{every edge}=[very thick]
\node (param)     at (-2,6)  {\parbox{6em}{(uncertain) \\ parameters}};
\node (bound)     at (2,6)  {\parbox{8em}{Forcing and boundary conditions}};
\node (gcm)       at (0,4)  {climate simulation};
\node (discrepancy) [fill=red!20]       at (-3.5,4)  {discrepancy};
\node (climate)       at (0,2)  {climate estimate};
\node (paramp)     at (-3,2)  {\parbox{6em}{(uncertain) \\ parameters}};
\node (proxysim)   at (0,0)  {Observation simulation};
\node (disproxy) [fill=red!20]  at (-3.5,0)  {discrepancy};
\node (proxyest) [fill=green!20]  at (0, -2)  {Observation};
\draw[->] (param) -> (gcm);
\draw[->] (bound) -> (gcm);
\draw[->] (gcm) -> (climate);
\draw[->] (discrepancy) -> (climate);
\draw[->] (climate) -> (proxysim);
\draw[->] (paramp) -> (proxysim);
\draw[->] (proxysim) -> (proxyest);
\draw[->] (disproxy) -> (proxyest);
\tikzstyle{every node}=[draw=none];
\end{tikzpicture}
\end{center}
\caption{Information provided by climate simulators may be involved in a hierarchical model similar to \figref{fig:statmodel}. Climate is modelled as the combination of climate simulations and discrepancy; and observations are modelled as the combination of a process-based observation model (for example: a vegetation model such as BIOME) and discrepancy. }
\label{fig:bayesiansim}
\end{figure}
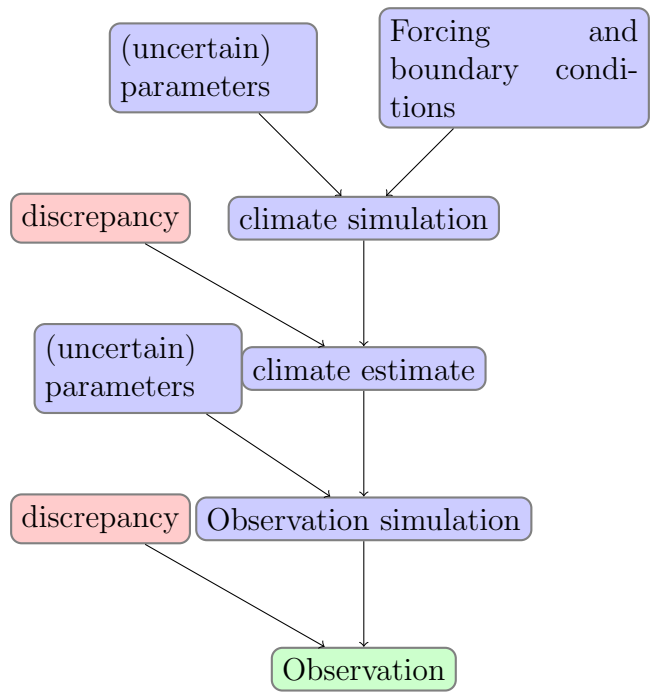

Unfortunately, process-based models are imperfect. Hence, the output of a process-based model cannot be taken as a plausible exact representation of reality, even if the model parameters are carefully chosen (\citet{Goldstein06aa} and \citet{Rougier2007Probabilistic-i}).
Consequently, it was argued that one cannot really be satisfied  with sampling model parameters in order to generate a prior distribution of plausible climates  \citet{Rougier2007Probabilistic-i}. Statisticians and climate scientists must therefore co-operate in order to develop a representation of the residual distance between the climate simulator and reality. This distance is called 
\textit{discrepancy} \citep{Goldstein06aa}. The discrepancy is a model in itself. It usually involves parameters, which can be defined a priori and updated.
For example, \cite{Schmittner11aa} use the combination of a constant bias and a spatially-covariate process to link the results of the UVic model to continental and marine palaeodata, although more work is still needed to quantify the effect of this discrepancy on current estimates of climate sensitivity.
These considerations lead us to a proposal programme for inference combining simulators with statistical reasoning that follows the schematics shown on \figref{fig:bayesiansim}. 
At least three challenges are being faced in this enterprise: sampling priors, expressing discrepancy, and identifying a tractable resolution algorithm. 
The present review is concerned with sampling priors only, because this is the area where progress has been the most tangible over the recent years. 

\subsection{Time slice modelling}

To illustrate the challenge consider the problem of reconstructing the climate of the Last Glacial Maximum. Suppose the climate-process model is a simulator like HadCM3 \citep{gordon00}, or any similar climate simulator. Experts are being elicited to identify a short-list of, say, 10 uncertain input parameters, which are thought to influence the simulated climate significantly. The problem is to sample this 10-dimension parameter space. If, for every parameter, 5 different values are being tried, $5^{10}$ experiments need to be run. That makes a bit less that 10 million experiments, each of which takes several weeks of super-computer time. This is naturally impossible.

On this question, the literature opposes two approaches.
The first one consists in searching methodically the region of maximum likelihood (or, maximum posterior) and estimate the local curvature of the likelihood (or, posterior) around it. 
To this end, the modeller may resort to the \textit{adjoint} of the simulator. This is a mathematical or software device that estimates sensitivity (formally: the gradient) of the model state  with respect to its parameters. The adjoint method was recently illustrated in the palaeoclimate context \citep{Paul12aa}. The reader is referred to textbooks for a more formal definition adjoint approach,  including \citet{Kalnay02aa} and \citet{Wunsch06aa}. The approach is efficient, but it may fail if the posterior has several local maxima. 

The second approach is to sample the 10-dimensional parameter space sparsely but smartly. The challenge is then to chose carefully the experiments to be run, such as to extract as much information as possible. This is a problem of `experiment design for computer experiments', for which a theory is being developed \citep{santner03}, and several articles focus explicitly on climate applications \citep{Urban10aa,Challenor11aa}. Once experiments have been run, the output of the simulator has to be interpolated throughout the entire 10-dimensional space. This reminds us of the spatial interpolation problem (interpolate sparse observations on a regular mesh). 
Unsurprisingly, interpolation methods for computer experiments use some of the mathematics of spatial modelling and, in particular, the Gaussian process \citep{Rasmussen06aa}. 
The literature on the subject mainly relies on the Bayesian formalism, and the interpolator is called in this context an `emulator' \citep{Kennedy00aa,Oakley02aa}. The modeller's choice may be to emulate global quantities, such as the globally averaged temperature \citep{Olson12aa}. In practice, though, palaeoclimate applications often require to reconstruct the spatial distribution of temperature, precipitation, and other climatic quantities. Several solutions have been proposed, including the Principal-Component-Analysis-emulator \citep{Wilkinson10aa} and the outer-product emulator \citep{Rougier08aa}.  Both options were tested in \citet{Schmittner11aa} (in their supplementary material).

\subsection{Dynamical modelling}
Let us  now focus on the dynamical climate evolution over Quaternary time scales. The role of the climate process-based model is to provide a physical prior on the physical relationships between the different climatic variables. 

Here, one should distinguish two different possible objectives. One objective is to \textit{reconstruct} a coherent picture of climate evolution, using physical constraints. This is a problem of \textit{state estimation}. Another objective is to enquire about climate mechanisms by testing hypotheses and estimate parameters. The second task can be translated into a Bayesian statistical framework as a problem of \textit{parameter estimation}. Here we review state and parameter estimation separately. 

\subsubsection{State estimation}
In principle state estimation in dynamical systems is a fairly standard problem for which a number of solutions exist, based on the Kalman filter \cite[e.g.][]{Evensen00aa} or the particle filter \citep{dfg01}. In practice, palaeoclimate applications involve numerical systems for which a direct application of theses technologies would be overwhelmingly expensive.  Consequently, somewhat ad hoc, application-dependent algorithms were suggested. 

For example, \cite{Bintanja05aa} use a thermodynamical ice-sheet model, including an explicit representation of oxygen isotope balance within the ice sheet, coupled to a simple representation of the  deep ocean and the atmosphere. Given observations of only certain variables of the model (here: deep ocean $\delta^{18}$O), the problem is to reconstruct the state of all system variables, and, in particular,  deep-ocean temperature and continental ice mass. This is a typical problem of state estimation, which was solved in this case by a time-step wise iterative procedure.

Another palaeoclimate application is the spatio-temporal reconstruction---for example, reconstructing the geographical distribution of temperature throughout the last millennium---using climate simulators. At the palaeoclimate time scale the only tractable approach today relies on low resolution general circulation models belonging to the category of climate models of intermediate complexity. If the ocean or atmosphere dynamics are chaotic (this is the case in `LOVECLIM', \citet{Goosse05aa}) then one needs to assimilate observations sequentially, that is, one after the next as time progresses. 
\citet{Dubinkina11aa} and \citet{Goosse12aa} have proposed to solve this problem with a low-computational version of the particle filter. 

\subsubsection{Parameter estimation}
Calibrating the parameters in a model on observations allows one to update prior knowledge on the relationships between different climate variables.
The most tractable solution (i.e., the least computational intensive) is to rely on a low-order dynamical system, for example a conceptual model, of which the execution is almost instantaneous on a computer. 
Even though the model is highly idealised, inference on model parameters may provide information on causal relationships between the different components of the Earth system. For example, if one parameter controls the effect of precession on ocean circulation, it is informative to know if this parameter is with high probability different to zero. 
One model proposed in the literature for such applications is the \citet{saltzman90sm} model. The methodological developments aiming at using this model in a statistical setting have paralleled those for snap-shot reconstructions discussed in the previous subsection: Early attempts are based on sampling uncertain parameters assuming that the model is perfect \citep{hargreaves02}, and later attempts have included a stochastic component to account for model discrepancy \citep{Crucifix09aa}. If the stochastic model is unstable, i.e., if two climate reconstructions simulated with two different but plausible realisations of the model error term vary widely, then the introduction of stochastic processes makes the parameter estimation task a particularly difficult problem, about which literature is scattered between the traditions of physics and dynamical system theory (see \citet{Voss04aa} for a general review, and \citet{Kwasniok09aa} for an application on Dansgaard-Oeschger events), and statistics (see \citet{Ionides06aa, lw01} and \cite{Andrieu10ab} for three different approaches to the problem).

\begin{figure}
\begin{center}
{\large\bf Dynamic emulator project }\\
\bigskip
\begin{tikzpicture}
\tikzstyle{every node}=[text centered, rounded corners, text width=6em, draw=black!50, thick]
\node (spar)     at (-5,2)  {simulator params};
\node (ic)     at (-2,2)  {initial conditions};
\node (icr)     at (2,2)  {initial conditions (reduced space)};
\node (tpar)     at (5,2)  {stoch. model params };
\node (climate)[sharp corners]     at (-4, -2)  {climate (space-time process)};
\node (sclimate)[sharp corners]    at (4, -2)  {climate (reduced space)};

\draw[->] (spar) -> (climate); 
\draw[->] (ic) -> (climate); 

\draw[->] (tpar) -> (sclimate); 
\draw[->] (icr) -> (sclimate); 
\draw[->] (ic) -> (icr); 

\path (sclimate) edge [bend right, ->, dashed]  (tpar) ;
\node (ra) [draw=none] at (7, 0)  {$\Rightarrow$};
\node (spar2)     at (10, 2)  {simulator params};
\node (tpar2)     at (10, -2)  {stoch. model params };
\path (climate) edge [bend angle=30, ->, dashed]  (sclimate) ;
\path (sclimate) edge [->, dashed]  (climate) ;

\draw[->] (spar2) -> (tpar2); 

\end{tikzpicture}
\end{center}
\caption{In the dynamic emulator project, for which \cite{Young11aa} provide theoretical background and applications,  the simulator is identified to a reduced, stochastic model, operating in a reduced space. For example, instead of the full geographical distribution, the stochastic dynamical system works on the space of principal components. For every set of parameters of the simulator, a corresponding set of parameters for the stochastic dynamical system can be estimated such as to reproduce results obtained with the full simulator as well as possible. The dynamical emulator then consists of the combination of the stochastic dynamical system with a statistical model, mapping simulator parameters on stochastic model parameters.}
\label{fig:dynemul}
\end{figure}
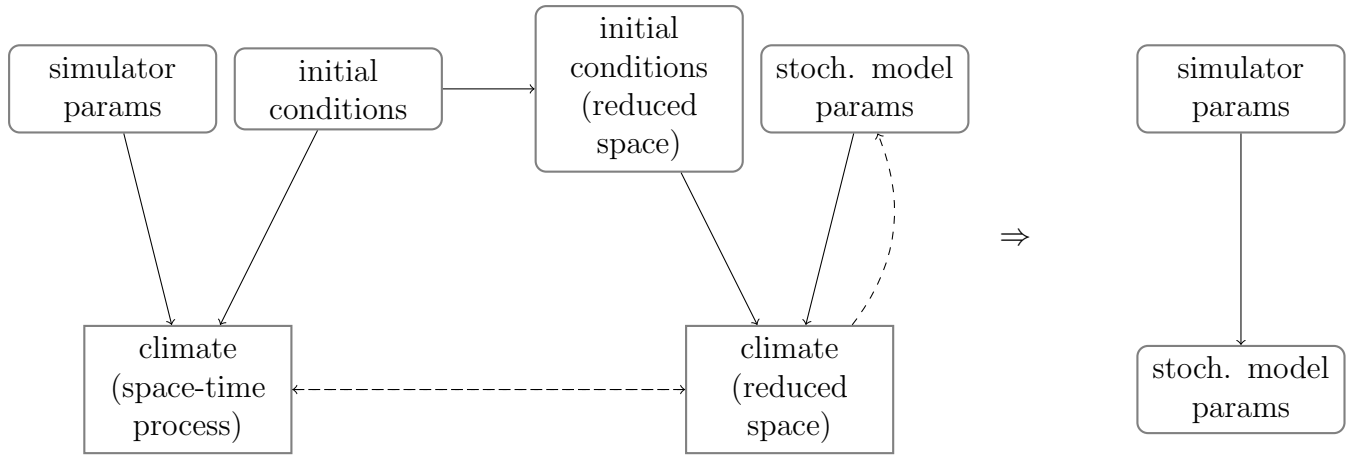

It may be objected that conceptual models contain too little physical information, and that using more comprehensive simulators is desirable.  The objection has merit, but how to deal with the subsequent intractable computational would unavoidably emerge?
%
%
\citet{Young11aa} proposed a strategy to address this specific problem, relying on the technique of dynamical system identification. This a two-step process \figrefp{fig:dynemul}. 

The first step consists in finding a stochastic dynamical system that reproduces adequately the ouptut of a simulator that  encapsulates the desired physics. A number of practical examples are already available in the literature:
\begin{description}
\item[1.] \cite{Hooss01aa} considered an ocean carbon cycle model developed to compute the oceanic uptake of CO$_2$ called HAMOCC1. HAMOCC1 is a complex simulator, resolving  the three-dimensional fluid dynamics and a number of oceanic tracers involved in the carbon and nutrient cycles. The model outputs three-dimensional fields. \cite{Hooss01aa} estimated the principal spatial patterns of the simulator output, and calibrated a finite impulse response model predicting the response of the coefficients of the principal patterns, in order to reproduce the full output of HAMOCC1. They  further corrected the model slightly to account for second-order non-linearities related to CO$_2$ uptake. 
\item[2.] The LLN2D model is a simulator of the growth and decay of the three major northern ice sheets, in response to variations in carbon dioxide and astronomical forcing \citep{gallee92}. The proposed surrogate of this model \citep{Crucifix11aa} is  a second-order (hence, non-linear) equation governing the total amount of continental ice, suitable to account for the co-existence of two stable steady-states in this model.
\item[3\&4.] Finally, \citet{Young09aa} considers, among others, two applications: (a) the \cite{Enting93aa} model of the global carbon cycle, which is itself a somewhat elaborate dynamical system,  with 23 coupled nonlinear equations and (b) the HadCM3 climate simulator \citep{gordon00}, used in response to anthropogenic greenhouse gas increases scenarios. For both applications a surrogate model is obtained by calibrating a `Transfer function model'  \footnote{
Transfer-function models are classical in dynamical system identification theory \citep{Ljung99aa}.
The assumption is that moving-averaged outputs and moving averaged-inputs are linearly correlated to each other. 
}
\end{description}                    
The second step consists of the \cite{Young11aa} strategy is to model the relationship between the parameters of the simulator and the parameters of the surrogate stochastic dynamical system using a Gaussian process. One then obtains a solution to explore the response of the simulator to different combinations of parameter and forcing, by actually running the surrogate stochastic dynamical system. In this context, \cite{Young11aa} calls the combination of the surrogate dynamical system with the model of the relationship between the parameters of the simulator and those of the surrogate system, a \textit{dynamical emulator}. So far no palaeoclimate application of this technique is available. 
\subsection{Emulator by-products}
The emulator was introduced and defined as a statistical surrogate of a simulator used to sample efficiently the space of uncertain parameters.  More generally, the whole process of designing and calibrating an emulator is a procedure aiming at extracting information efficiently from a complex numerical system. This is a computing cost economiser, which comes with at least two possible applications in addition to the problems of parameter and state estimation discussed above:
\begin{enumerate}
\item Once the emulator is properly calibrated and tested, there is almost no limit to the number of sensitivity experiments to model inputs that may be run, because they are run with the emulator rather than with the actual simulator. Consequently, the sensitivity of the simulator can be explored very finely and span the whole input space, which includes boundary conditions, forcing and parameters. The procedure provides us with a way to characterise the non-linear structure of the model response and, in particular, identify thresholds and plateaus.  \cite{Oakley04aa} provide some theoretical background to this end. 
\item Given the computational resources required to run general circulation models in palaeoclimate applications, modellers generally chose to degrade the spatial resolution and simplify the physics (for example: assume that cloud cover is constant, or simplify the relationship between large scale atmospheric dynamics and precipitation) compared to  state of the art simulators. This is the rationale for the use of Earth System Models of Intermediate Complexity \citep{claussen02emic}. However, these physical simplifications have deleterious effects on the representation and the response structure of certain processes thought to play a role at the global scale, including tropical monsoons, El-Ni\~no, Southern ocean ventilation and North Atlantic convection. The emulator provides an alternative to this reduced-physics strategy, by offering the possibility of running numerous experiments with an emulator calibrated on a simulator, rather than with the simulator itself.
\end{enumerate}
This being said, developing an emulator is by no means straightforward, be it in the static or in the dynamic framework. The choice of emulator structure,  inputs, and regressors is a fairly heuristic procedure that requires expertise on the simulator and the processes that it is meant to represent. 
In particular, one cannot overstate the importance of extensive validation tests of the emulator: the user must be confident that the emulator is consistent with the simulator dynamics and avoid unsuitable interpretations. The literature on emulator validation and design is developing \citep{Bastos09aa} (cf. also the \url{www.mucm.ac.uk} website) and should be carefully examined before any application.
The natural recommendation, echoed by \cite{Young11aa} (and references therein) is to adopt a framework that is as simple as possible, chosen such that the parameters of the surrogate model can be interpreted in terms of system dynamics and system sensitivity. 

\section{Conclusion}
The present article reviewed three modelling approaches adequate to study palaeoclimate phenomena:
\begin{enumerate}
\item Conceptual models provide a  representation of a well-identified phenomenon like ice ages, Dansgaard-Oeschger events or  Heinrich events. These models are specific to the studied phenomenon, and their development obeys a principle of parsimony: explaining as much as possible with as little as possible. In spite of this principle of parsimony, there may be a risk of over-interpretation. Two conceptual models relying on different mechanistic interpretations may have similar mathematical descriptions, and thus fit the same dataset equally well \citep{tziperman06pacing,Cane:2006,Crucifix12aa}. Consequently, a model fit does not necessarily imply that the  underlying causal interpretation is correct. As a general rule, the model will be  more convincing if it accommodates different forms of information, that is: different records ($\delta^{18}O$, CO$_2$, SST,\ldots), different aspects of the record (trends, noise, auto-correlation,\ldots), and captures transitions between different regimes (for example: the Middle Pleistocene Revolution) with few free parameters. The theory of dynamical system tells us that chaotic and other counter-intuitive mathematical phenomena may arise in deceptively simple systems. For this reason, dynamical system identification, which we briefly reviewed at the end of the article, is useful as a tool for analysing the dynamics of  simulators used to generate time series. If the behaviour of a complex simulator can be reproduced with a compact, stochastic dynamical system, it can be more easily analysed from the prospective of dynamical system theory. 
\item Simulators provide a spatially-resolved representation of numerous physical and biogeochemical processes involved in the climate system. Simulators are more versatile in nature than conceptual models. Depending on the experimental design, the same climate simulator may be used to study different climatic phenomena at different time and/or spatial scales. Causal effects may be explored by means of intervention experiments, in which one of the components of the system is being taken control of (e.g.: fixed or `imposed' ice sheets, vegetation, CO$_2$ etc.) However, it would be a mistake to regard simulators as \textit{ab initio} models of the climate system. To the extent simulators and their sub-grid scale parameterisations are tuned to adequately reproduce planetary scale phenomena (globally-average temperature, the Atlantic meridional overturning cell,  monsoon dynamics), simulators are partly empirical. Scientists working with climate simulators are aware of this and for this reason they develop stringent tests to assess the performance of these models, both for the representation of the present-day climate and for the past climates. 
\item Statistical modelling accommodates systematic relationships---but not necessarily causal---between different variables with random effects. Statistical modelling may in particular be involved in space-time climate reconstruction, hypothesis testing and parameter estimation, but it should never be overlooked that these mathematical operations are constrained by a number of hypotheses expressed in the statistical model. Common statistical assumptions concern the autocorrelation of data in time and space, and between the climate state and actual observations. Hierarchical modelling is increasingly used because it provides a more explicit representation of the forward mechanical process linking the causes of the phenomenon to actual observations.
\end{enumerate}

One of the purposes of the review is to show that statistics, and Bayesian statistics in particular, have a role to play to articulate these frameworks. In particular, they allow us to include information on a priori physically plausible model parameter ranges in the inference process. 

Unfortunately there is no free lunch. 
A brute-force sampling approach may involve  millions or even billions of experiments, which cannot be envisaged even with the simplest systems. Consequently, resolution algorithms tend to be mathematically highly involved, and resort to intermediate modelling steps such as the emulator or the adjoint, experiment planning, along with new and not yet solidly established concepts such as the discrepancy. These modelling steps and associated resolution algorithms need to be evaluated carefully and this procedure is tedious and technical. 
What has been gained on the one hand (a more transparent expression of the conditions of inference) has to be paid by the technical character of the inference process, which may appear obscure and off-putting to a significant fraction of the scientific audience. 

The conclusion follows naturally. The established paradigms of modelling that were reviewed here (conceptual, simulator and statistical modelling) still have an important role to play because they allow scientists to build on years of achievement and experience. Conceptual modelling, in particular, is a flexible approach to organise observations involved in complex climatic processes. On the other hand, the power of cross-paradigm approaches, allying statistics with simulators, or statistics with conceptual models, should not be overlooked. They offer avenues to manage uncertainties and rationalise model design and selection given different and variate sources of information.

\section*{Acknowledgements} 
I dedicate this article to the many authors of the admirable articles  from the 70s and 80s that were cited here. They are an endless source of inspiration and motivation.
The review benefited from interactions with Jonathan Rougier (University of Bristol), Didier Paillard (LSCE, Gif-sur-Yvette), Nabila Bounceur and Carlos Almeida (both from Universit\'e catholique de Louvain), as well as from the comments by two reviewers. I am also indebted to Caitlin Buck (University of Sheffield) for managing the SUPRAnet international scientific network (\url{http://caitlin-buck.staff.shef.ac.uk/SUPRAnet/}), of which I have largely benefited.  I am research associate with the Belgian National Fund of Scientific Research, and this research is funded by the European Research Council, Starting Grant contract nr 2009-StG-239604 ITOP. 
\bibliography{/Users/crucifix/Documents/BibDesk.bib}

\begin{thebibliography}{155}
\providecommand{\natexlab}[1]{#1}
\expandafter\ifx\csname urlstyle\endcsname\relax
  \providecommand{\doi}[1]{doi:\discretionary{}{}{}#1}\else
  \providecommand{\doi}{doi:\discretionary{}{}{}\begingroup
  \urlstyle{rm}\Url}\fi

\bibitem[{\textit{Andrieu et~al.}(2010)\textit{Andrieu, Doucet, and
  Holenstein}}]{Andrieu10ab}
Andrieu, C., A.~Doucet, and R.~Holenstein, Particle {M}arkov chain {M}onte
  {C}arlo methods, \textit{Journal of the Royal Statistical Society: Series B
  (Statistical Methodology)}, \textit{72}(3), 269--342,
  \doi{10.1111/j.1467-9868.2009.00736.x}, 2010.

\bibitem[{\textit{Annan et~al.}(2005)\textit{Annan, Hargreaves, Ohgaito,
  Abe-Ouchi, and Emori}}]{annan05sola}
Annan, J.~D., J.~C. Hargreaves, R.~Ohgaito, A.~Abe-Ouchi, and S.~Emori,
  Efficiently constraining climate sensitivity with ensembles of paleoclimate
  experiments, \textit{SOLA}, \textit{1}, 181--184,
  \doi{10.2151/sola.2005-047}, 2005.

\bibitem[{\textit{Archer et~al.}(2000)\textit{Archer, Winguth, Lea, and
  Mahowald}}]{Archer00aa}
Archer, D., A.~Winguth, D.~Lea, and N.~Mahowald, What caused the
  glacial/interglacial atmospheric p{CO$_2$} cycles?, \textit{Reviews of
  Geophysics}, \textit{38}(2), 159--189, \doi{10.1029/1999RG000066}, 2000.

\bibitem[{\textit{Bastos and O'Hagan}(2009)}]{Bastos09aa}
Bastos, L.~S., and A.~O'Hagan, Diagnostics for {G}aussian process emulators,
  \textit{Technometrics}, \textit{51}(4), 425--438,
  \doi{10.1198/TECH.2009.08019}, 2009.

\bibitem[{\textit{Benzi et~al.}(1982)\textit{Benzi, Parisi, Sutera, and
  Vulpiani}}]{BENZI82aa}
Benzi, R., G.~Parisi, A.~Sutera, and A.~Vulpiani, Stochastic resonance in
  climatic change, \textit{Tellus}, \textit{34}(1), 10--16,
  \doi{10.1111/j.2153-3490.1982.tb01787.x}, 1982.

\bibitem[{\textit{Berger}(1978)}]{berger78}
Berger, A.~L., Long-term variations of daily insolation and {Q}uaternary
  climatic changes, \textit{J. Atmos. Sci.}, \textit{35}, 2362--2367,
  \doi{10.1175/1520-0469(1978)035<2362:LTVODI>2.0.CO;2}, 1978.

\bibitem[{\textit{Bintanja et~al.}(2005)\textit{Bintanja, van~de Wal, and
  Oerlemans}}]{Bintanja05aa}
Bintanja, R., R.~S.~W. van~de Wal, and J.~Oerlemans, Modelled atmospheric
  temperatures and global sea levels over the past million years,
  \textit{Nature}, \textit{437}(7055), 125--128, \doi{10.1038/nature03975},
  2005.

\bibitem[{\textit{Blaauw and Christen}(2005)}]{Blaauw05aa}
Blaauw, M., and J.~A. Christen, Radiocarbon peat chronologies and environmental
  change, \textit{Journal of the Royal Statistical Society: Series C (Applied
  Statistics)}, \textit{54}(4), 805--816,
  \doi{10.1111/j.1467-9876.2005.00516.x}, 2005.

\bibitem[{\textit{Braconnot et~al.}(2012)\textit{Braconnot, Harrison, Kageyama,
  Bartlein, Masson-Delmotte, Abe-Ouchi, Otto-Bliesner, and
  Zhao}}]{Braconnot12aa}
Braconnot, P., S.~P. Harrison, M.~Kageyama, P.~J. Bartlein, V.~Masson-Delmotte,
  A.~Abe-Ouchi, B.~Otto-Bliesner, and Y.~Zhao, Evaluation of climate models
  using palaeoclimatic data, \textit{Nature Climate Change}, \textit{2}(6),
  417--424, \doi{10.1038/nclimate1456}, 2012.

\bibitem[{\textit{Braconnot et~al.}(2007)}]{braconnot07pmip1}
Braconnot, P., et~al., Results of {P}{M}{I}{P}2 coupled simulations of the
  {M}id {H}olocene and {L}ast {G}lacial {M}aximum -- {P}art 1: experiments and
  large-scale features, \textit{Climate of the Past}, \textit{3}, 261--277,
  \doi{10.5194/cp-3-261-2007}, 2007.

\bibitem[{\textit{Braun et~al.}(2007)\textit{Braun, Ganopolski, Christl, and
  Chialvo}}]{Braun07aa}
Braun, H., A.~Ganopolski, M.~Christl, and D.~R. Chialvo, A simple conceptual
  model of abrupt glacial climate events, \textit{Nonlinear Processes in
  Geophysics}, \textit{14}(6), 709--721, \doi{10.5194/npg-14-709-2007}, 2007.

\bibitem[{\textit{Braun et~al.}(2011)\textit{Braun, Ditlevsen, Kurths, and
  Mudelsee}}]{Braun11aa}
Braun, H., P.~Ditlevsen, J.~Kurths, and M.~Mudelsee, A two-parameter stochastic
  process for {D}ansgaard-{O}eschger events, \textit{Paleoceanography},
  \textit{26}(3), PA3214, \doi{10.1029/2011PA002140}, 2011.

\bibitem[{\textit{Brewer et~al.}(2007)\textit{Brewer, Guiot, and
  Torre}}]{Brewer07aa}
Brewer, S., J.~Guiot, and F.~Torre, Mid-{H}olocene climate change in {E}urope:
  a data-model comparison, \textit{Climate of the Past}, \textit{3}(3),
  499--512, \doi{10.5194/cp-3-499-2007}, 2007.

\bibitem[{\textit{Broecker}(1991)}]{Broecker91aa}
Broecker, W.~S., The great ocean conveyor, \textit{Oceanography}, \textit{4},
  79--89, 1991.

\bibitem[{\textit{Brovkin et~al.}(1998)\textit{Brovkin, Claussen, Petoukhov,
  and Ganopolski}}]{Brovkin98aa}
Brovkin, V., M.~Claussen, V.~Petoukhov, and A.~Ganopolski, On the stability of
  the atmosphere-vegetation system in the sahara/sahel region, \textit{J.
  Geophys. Res.}, \textit{103}(D24), 31,613--31,624,
  \doi{10.1029/1998JD200006}, 1998.

\bibitem[{\textit{Brovkin et~al.}(2003)\textit{Brovkin, Levis, Loutre,
  Crucifix, Claussen, Ganopolski, Kubatzki, and Petoukhov}}]{brovkin03cc}
Brovkin, V., S.~Levis, M.~F. Loutre, M.~Crucifix, M.~Claussen, A.~Ganopolski,
  C.~Kubatzki, and V.~Petoukhov, Stability analysis of the climate-vegetation
  system in the northern high latitudes, \textit{Clim. Change},
  \textit{57}(1-2), 119--138, 2003.

\bibitem[{\textit{Brovkin et~al.}(2012)\textit{Brovkin, Ganopolski, Archer, and
  Munhoven}}]{Brovkin12aa}
Brovkin, V., A.~Ganopolski, D.~Archer, and G.~Munhoven, Glacial {CO}$_{2}$
  cycle as a succession of key physical and biogeochemical processes,
  \textit{Climate of the Past}, \textit{8}(1), 251--264,
  \doi{10.5194/cp-8-251-2012}, 2012.

\bibitem[{\textit{Calov et~al.}(2005)\textit{Calov, Ganopolski, Claussen,
  Petoukhov, and Greve}}]{Calov2005Transient-simul}
Calov, R., A.~Ganopolski, M.~Claussen, V.~Petoukhov, and R.~Greve, Transient
  simulation of the last glacial inception. {P}art {I}: glacial inception as a
  bifurcation in the climate system, \textit{Climate Dynamics}, \textit{24}(6),
  545--561, \doi{10.1007/s00382-005-0007-6}, 2005.

\bibitem[{\textit{Cane et~al.}(2006)}]{Cane:2006}
Cane, M.~A., et~al., Progress in paleoclimate modeling, \textit{Journal of
  Climate}, \textit{19}(20), 5031--5057, \doi{10.1175/JCLI3899.1}, 2006.

\bibitem[{\textit{Challenor}(2011)}]{Challenor11aa}
Challenor, P., Designing a computer experiment that involves switches,
  \textit{Journal of Statistical Theory and Practice}, \textit{5}(1), 47--57,
  \doi{10.1080/15598608.2011.10412049}, 2011.

\bibitem[{\textit{Chikamoto et~al.}(2012)\textit{Chikamoto, Abe-Ouchi, Oka,
  Ohgaito, and Timmermann}}]{Chikamoto12aa}
Chikamoto, M.~O., A.~Abe-Ouchi, A.~Oka, R.~Ohgaito, and A.~Timmermann,
  Quantifying the ocean's role in glacial {CO}$_{2}$ reductions,
  \textit{Climate of the Past}, \textit{8}(2), 545--563,
  \doi{10.5194/cp-8-545-2012}, 2012.

\bibitem[{\textit{Claussen et~al.}(1999)\textit{Claussen, Kubatzki, Brovkin,
  Ganopolski, Hoelzmann, and Pachur}}]{claussen99}
Claussen, M., C.~Kubatzki, V.~Brovkin, A.~Ganopolski, P.~Hoelzmann, and H.-J.
  Pachur, Simulation of an abrupt change in {Saharan} vegetation in the
  mid-{H}olocene, \textit{Geophysical Research Letters}, \textit{26},
  2037--2040, \doi{10.1029/1999GL900494,}, 1999.

\bibitem[{\textit{Claussen et~al.}(2002)}]{claussen02emic}
Claussen, M., et~al., Earth system models of intermediate complexity: closing
  the gap in the spectrum of climate system models, \textit{Climate Dynamics},
  \textit{18}(7), 579--586, \doi{10.1007/s00382-001-0200-1}, 2002.

\bibitem[{\textit{Crucifix}(2011)}]{Crucifix11aa}
Crucifix, M., How can a glacial inception be predicted?, \textit{The Holocene},
  \textit{21}(5), 831--842, \doi{10.1177/0959683610394883}, 2011.

\bibitem[{\textit{Crucifix}(2012)}]{Crucifix12aa}
Crucifix, M., Oscillators and relaxation phenomena in pleistocene climate
  theory, \textit{Philosophical Transactions of the Royal Society A:
  Mathematical, Physical and Engineering Sciences}, \textit{370}(1962),
  1140--1165, \doi{10.1098/rsta.2011.0315}, 2012.

\bibitem[{\textit{Crucifix and Rougier}(2009)}]{Crucifix09aa}
Crucifix, M., and J.~Rougier, On the use of simple dynamical systems for
  climate predictions: A {B}ayesian prediction of the next glacial inception,
  \textit{European Physics Journal - Special Topics}, \textit{174}, 11--31,
  \doi{10.1140/epjst/e2009-01087-5}, 2009.

\bibitem[{\textit{Dakos et~al.}(2008)\textit{Dakos, Scheffer, van Nes, Brovkin,
  Petoukhov, and Held}}]{Dakos08aa}
Dakos, V., M.~Scheffer, E.~H. van Nes, V.~Brovkin, V.~Petoukhov, and H.~Held,
  Slowing down as an early warning signal for abrupt climate change,
  \textit{Proceedings of the National Academy of Sciences of the United States
  of America}, \textit{105}(38), 14,308--14,312, \doi{10.1073/pnas.0802430105},
  2008.

\bibitem[{\textit{Davison}(2003)}]{Davison03aa}
Davison, A.~C., \textit{Statistical models}, Cambridge Series in Statistical
  and Probabilistic Modelling, {Cambridge University Press}, 2003.

\bibitem[{\textit{De~Saedeleer et~al.}(2012)\textit{De~Saedeleer, Crucifix, and
  Wieczorek}}]{De-Saedeleer12aa}
De~Saedeleer, B., M.~Crucifix, and S.~Wieczorek, Is the astronomical forcing a
  reliable and unique pacemaker for climate? a conceptual model study,
  \textit{Climate Dynamics}, p. online first, \doi{10.1007/s00382-012-1316-1},
  2012.

\bibitem[{\textit{Ditlevsen}(2009)}]{Ditlevsen09aa}
Ditlevsen, P.~D., Bifurcation structure and noise-assisted transitions in the
  {P}leistocene glacial cycles, \textit{Paleoceanography}, \textit{24}, PA3204,
  \doi{10.1029/2008PA001673}, 2009.

\bibitem[{\textit{Ditlevsen and Johnsen}(2010)}]{Ditlevsen10aa}
Ditlevsen, P.~D., and S.~J. Johnsen, Tipping points: Early warning and wishful
  thinking, \textit{Geophys. Res. Lett.}, \textit{37}(19), L19,703,
  \doi{10.1029/2010GL044486}, 2010.

\bibitem[{\textit{Doucet et~al.}(2001)\textit{Doucet, de~Freitas, and
  Gordon}}]{dfg01}
Doucet, A., N.~de~Freitas, and N.~Gordon (Eds.), \textit{Sequential {M}onte
  {C}arlo Methods in Practice}, New York: Springer, 2001.

\bibitem[{\textit{Dubinkina et~al.}(2011)\textit{Dubinkina, Goosse,
  Sallaz-Damaz, Crespin, and Crucifix}}]{Dubinkina11aa}
Dubinkina, S., H.~Goosse, Y.~Sallaz-Damaz, E.~Crespin, and M.~Crucifix, Testing
  a particle filter to reconstruct climate changes over the past centuries,
  \textit{International Journal of Bifurcation and Chaos}, \textit{21},
  3611--3618, \doi{10.1142/S0218127411030763}, 2011.

\bibitem[{\textit{Eckermann}(2011)}]{Eckermann11aa}
Eckermann, S.~D., Explicitly stochastic parameterization of nonorographic
  gravity wave drag, \textit{Journal of the Atmospheric Sciences},
  \textit{68}(8), 1749--1765, \doi{10.1175/2011JAS3684.1}, 2011.

\bibitem[{\textit{Enting and Lassey}(1993)}]{Enting93aa}
Enting, I.~G., and K.~R. Lassey, Projections of future {CO}$_2$.,
  \textit{Technical Report~27}, Division of Atmoshperic Research, CSIRO,
  Melbourne, 1993.

\bibitem[{\textit{Evensen and van Leeuwen}(2000)}]{Evensen00aa}
Evensen, G., and P.~J. van Leeuwen, An ensemble {K}alman smoother for nonlinear
  dynamics, \textit{Monthly Weather Review}, \textit{128}(6), 1852--1867,
  \doi{10.1175/1520-0493(2000)128<1852:AEKSFN>2.0.CO;2}, 2000.

\bibitem[{\textit{Gall{\'e}e et~al.}(1992)\textit{Gall{\'e}e, van Ypersele,
  Fichefet, Marsiat, Tricot, and Berger}}]{gallee92}
Gall{\'e}e, H., J.~P. van Ypersele, T.~Fichefet, I.~Marsiat, C.~Tricot, and
  A.~Berger, Simulation of the last glacial cycle by a coupled, sectorially
  averaged climate-ice sheet model. {Part II} : Response to insolation and
  {CO}$_2$ variation, \textit{Journal of Geophysical Research}, \textit{97},
  15,713--15,740, \doi{10.1029/92JD01256}, 1992.

\bibitem[{\textit{Ganopolski and Rahmstorf}(2002)}]{Ganopolski2002prl}
Ganopolski, A., and S.~Rahmstorf, Abrupt glacial climate changes due to
  stochastic resonance, \textit{Phys. Rev. Lett.}, \textit{88}(3), 038,501,
  \doi{10.1103/PhysRevLett.88.038501}, 2002.

\bibitem[{\textit{Ganopolski and Roche}(2009)}]{Ganopolski09aa}
Ganopolski, A., and D.~M. Roche, On the nature of lead--lag relationships
  during glacial--interglacial climate transitions, \textit{Quaternary Science
  Reviews}, \textit{28}(27--28), 3361--3378,
  \doi{10.1016/j.quascirev.2009.09.019}, 2009.

\bibitem[{\textit{Ghil}(1976)}]{Ghil76aa}
Ghil, M., Climate stability for a sellers-type model, \textit{Journal of the
  Atmospheric Sciences}, \textit{33}(1), 3--20, 1976.

\bibitem[{\textit{Ghil and Ghildress}(1987)}]{Ghil87aa}
Ghil, M., and S.~Ghildress, \textit{Topics in Geophysical Fluid Dynamics:
  Atmospheric Dynamics, Dynamo Theory and Climate Dynamics}, \textit{Applied
  Mathematical Sciences}, vol.~60, Springer-Verlag, 1987.

\bibitem[{\textit{Ghil and Le~Treut}(1981)}]{ghil81}
Ghil, M., and H.~Le~Treut, A climate model with cryodynamics and geodynamics,
  \textit{Journal of Geophysical Research}, \textit{86}(C6), 5262--5270,
  \doi{10.1029/JC086iC06p05262}, 1981.

\bibitem[{\textit{Goldstein and Rougier}(2006)}]{Goldstein06aa}
Goldstein, M., and J.~Rougier, Bayes linear calibrated prediction for complex
  systems, \textit{Journal of the American Statistical Association},
  \textit{101}, 1132--1143, \doi{10.1198/016214506000000203}, 2006.

\bibitem[{\textit{Goosse et~al.}(2005)\textit{Goosse, Renssen, Timmermann, and
  Bradley}}]{Goosse05aa}
Goosse, H., H.~Renssen, A.~Timmermann, and R.~S. Bradley, Internal and forced
  climate variability during the last millennium: a model-data comparison using
  ensemble simulations, \textit{Quaternary Science Reviews},
  \textit{24}(12--13), 1345--1360, \doi{10.1016/j.quascirev.2004.12.009}, 2005.

\bibitem[{\textit{Goosse et~al.}(2012)\textit{Goosse, Crespin, Dubinkina,
  Loutre, Mann, Renssen, Sallaz-Damaz, and Shindell}}]{Goosse12aa}
Goosse, H., E.~Crespin, S.~Dubinkina, M.-F. Loutre, M.~Mann, H.~Renssen,
  Y.~Sallaz-Damaz, and D.~Shindell, The role of forcing and internal dynamics
  in explaining the ``medieval climate anomaly'', \textit{Climate Dynamics}, p.
  online first, \doi{10.1007/s00382-012-1297-0}, 2012.

\bibitem[{\textit{Goosse et~al.}(2010)}]{Goosse10aa}
Goosse, H., et~al., Description of the {E}arth system model of intermediate
  complexity {LOVECLIM} version 1.2, \textit{Geoscientific Model Development},
  \textit{3}(2), 603--633, \doi{10.5194/gmd-3-603-2010}, 2010.

\bibitem[{\textit{Gordon et~al.}(2000)\textit{Gordon, Cooper, Senior, Banks,
  Gregory, Johns, Mitchell, and Wood}}]{gordon00}
Gordon, C., C.~Cooper, C.~A. Senior, H.~Banks, J.~M. Gregory, T.~C. Johns,
  J.~F.~B. Mitchell, and R.~A. Wood, The simulation of {SST}, sea ice extents
  and ocean heat transports in a version of the {H}adley {C}entre coupled model
  without flux adjustments, \textit{Climate Dynamics}, \textit{16}, 147--168,
  \doi{10.1007/s003820050010}, 2000.

\bibitem[{\textit{Gregory et~al.}(2004)\textit{Gregory, Ingram, Palmer, Jones,
  Stott, Thorpe, Lowe, Johns, and Williams}}]{Gregory04ab}
Gregory, J.~M., W.~J. Ingram, M.~A. Palmer, G.~S. Jones, P.~A. Stott, R.~B.
  Thorpe, J.~A. Lowe, T.~C. Johns, and K.~D. Williams, A new method for
  diagnosing radiative forcing and climate sensitivity, \textit{Geophysical
  Research Letters}, \textit{31}(3), L03,205, \doi{10.1029/2003GL018747}, 2004.

\bibitem[{\textit{Guillemot}(2010)}]{Guillemot10ab}
Guillemot, H., Connections between simulations and observation in climate
  computer modeling. scientist's practices and ``bottom-up epistemology"
  lessons, \textit{Studies In History and Philosophy of Science Part B: Studies
  In History and Philosophy of Modern Physics}, \textit{41}(3), 242--252,
  \doi{10.1016/j.shpsb.2010.07.003}, 2010.

\bibitem[{\textit{Guiot et~al.}(1999)\textit{Guiot, Boreux, Braconnot, and
  Torre}}]{Guiot99aa}
Guiot, J., J.~J. Boreux, P.~Braconnot, and F.~Torre, Data-model comparison
  using fuzzy logic in paleoclimatology, \textit{Climate Dynamics},
  \textit{15}(8), 569--581, \doi{10.1007/s003820050301}, 1999.

\bibitem[{\textit{Guiot et~al.}(2000)\textit{Guiot, Torre, Jolly, Peyron,
  Boreux, and Cheddadi}}]{Guiot2000Inverse-vegetat}
Guiot, J., F.~Torre, D.~Jolly, O.~Peyron, J.~J. Boreux, and R.~Cheddadi,
  Inverse vegetation modeling by monte carlo sampling to reconstruct
  palaeoclimates under changes precipitation seasonality and co$_2$ conditions:
  application to glacial climate in mediterranean region, \textit{Ecological
  Modelling}, \textit{127}, 119--140, \doi{10.1016/S0304-3800(99)00219-7},
  2000.

\bibitem[{\textit{Haam and Huybers}(2010)}]{Haam10aa}
Haam, E., and P.~Huybers, A test for the presence of covariance between
  time-uncertain series of data with application to the {D}ongge {C}ave
  speleothem and atmospheric radiocarbon records, \textit{Paleoceanography},
  \textit{25}(2), PA2209, \doi{10.1029/2008PA001713}, 2010.

\bibitem[{\textit{Hansen et~al.}(1984)\textit{Hansen, Lacis, Rind, Russell,
  Stone, Fung, Ruedy, and Lerner}}]{hansen84}
Hansen, J.~E., A.~Lacis, D.~Rind, L.~Russell, P.~Stone, I.~Fung, R.~Ruedy, and
  J.~Lerner, Climate sensitivity: analysis of feedback mechanisms, in
  \textit{Climate Processes and Climate Sensitivity}, edited by J.~E. Hansen
  and T.~Takahashi, pp. 130--163, American Geophysical Union, Washington D. C.,
  1984.

\bibitem[{\textit{Hargreaves and Annan}(2002)}]{hargreaves02}
Hargreaves, J.~C., and J.~D. Annan, Assimilation of paleo-data in a simple
  {E}arth system model, \textit{Climate Dynamics}, \textit{19}, 371--381,
  \doi{10.1007/s00382-002-0241-0}, 2002.

\bibitem[{\textit{Hargreaves et~al.}(2011)\textit{Hargreaves, Paul, Ohgaito,
  Abe-Ouchi, and Annan}}]{Hargreaves11aa}
Hargreaves, J.~C., A.~Paul, R.~Ohgaito, A.~Abe-Ouchi, and J.~D. Annan, Are
  paleoclimate model ensembles consistent with the {MARGO} data synthesis?,
  \textit{Climate of the Past}, \textit{7}(3), 917--933,
  \doi{10.5194/cp-7-917-2011}, 2011.

\bibitem[{\textit{Harrison and Prentice}(2003)}]{Harrison03lgm}
Harrison, S.~P., and I.~C. Prentice, Climate and {C}{O}$_{2}$ controls on
  global vegetation distribution at the {L}ast {G}lacial {M}aximum: analysis
  based on palaeovegetation data, biome modelling and palaeoclimate
  simulations, \textit{Global Change Biology}, \textit{9}(7), 983--1004,
  \doi{10.1046/j.1365-2486.2003.00640.x}, 2003.

\bibitem[{\textit{Haslett and Parnell}(2008)}]{Haslett08aa}
Haslett, J., and A.~Parnell, A simple monotone process with application to
  radiocarbon-dated depth chronologies, \textit{Journal of the Royal
  Statistical Society: Series C (Applied Statistics)}, \textit{57}(4),
  399--418, \doi{10.1111/j.1467-9876.2008.00623.x}, 2008.

\bibitem[{\textit{Haslett et~al.}(2006)\textit{Haslett, Whiley, Bhattacharya,
  Salter-Townshend, Wilson, Allen, Huntley, and Mitchell}}]{Haslett06aa}
Haslett, J., M.~Whiley, S.~Bhattacharya, M.~Salter-Townshend, S.~P. Wilson,
  J.~R.~M. Allen, B.~Huntley, and F.~J.~G. Mitchell, Bayesian palaeoclimate
  reconstruction, \textit{Journal of the Royal Statistical Society: Series A
  (Statistics in Society)}, \textit{169}(3), 395--438, 2006.

\bibitem[{\textit{Hasselmann}(1976)}]{Hasselmann76}
Hasselmann, K., Stochastic climate models part {I}. {T}heory, \textit{Tellus},
  \textit{28}(6), 473--485, \doi{10.1111/j.2153-3490.1976.tb00696.x}, 1976.

\bibitem[{\textit{Haxeltine and Prentice}(1996)}]{Haxeltine96aa}
Haxeltine, A., and I.~C. Prentice, {BIOME3}: An equilibrium terrestrial
  biosphere model based on ecophysiological constraints, resource availability,
  and competition among plant functional types, \textit{Global Biogeochem.
  Cycles}, \textit{10}(4), 693--709, \doi{10.1029/96GB02344}, 1996.

\bibitem[{\textit{Hays et~al.}(1976)\textit{Hays, Imbrie, and
  Shackleton}}]{hays76}
Hays, J.~D., J.~Imbrie, and N.~J. Shackleton, Variations in the {E}arth's orbit
  : {P}acemaker of ice ages, \textit{Science}, \textit{194}, 1121--1132,
  \doi{10.1126/science.194.4270.1121}, 1976.

\bibitem[{\textit{Held}(2005)}]{held05hierarchy}
Held, I.~M., The gap between simulation and understanding in climate modelling,
  \textit{Bull. Am. Meteorol. Soc.}, \textit{86}(11), 1609--1614,
  \doi{10.1175/BAMS-86-11-1609}, 2005.

\bibitem[{\textit{Hoffmann et~al.}(1998)\textit{Hoffmann, Werner, and
  Heimann}}]{Hoffmann98aa}
Hoffmann, G., M.~Werner, and M.~Heimann, Water isotope module of the {ECHAM}
  atmospheric general circulation model: A study on timescales from days to
  several years, \textit{Journal of Geophysical Research}, \textit{103}(D14),
  16,871--16,896, \doi{10.1029/98JD00423}, 1998.

\bibitem[{\textit{Hooss et~al.}(2001)\textit{Hooss, Voss, Hasselmann,
  Maier-Reimer, and Joos}}]{Hooss01aa}
Hooss, G., R.~Voss, K.~Hasselmann, E.~Maier-Reimer, and F.~Joos, A nonlinear
  impulse response model of the coupled carbon cycle-climate system ({NICCS}),
  \textit{Climate Dynamics}, \textit{18}, 189--202,
  \doi{10.1007/s003820100170}, 10.1007/s003820100170, 2001.

\bibitem[{\textit{Huybers and Wunsch}(2005)}]{huybers05obliquity}
Huybers, P., and C.~Wunsch, Obliquity pacing of the late {P}leistocene glacial
  terminations, \textit{Nature}, \textit{434}, 491--494,
  \doi{10.1038/nature03401}, 2005.

\bibitem[{\textit{Imbrie and Imbrie}(1980)}]{imbrie80}
Imbrie, J., and J.~Z. Imbrie, Modelling the climatic response to orbital
  variations, \textit{Science}, \textit{207}, 943--953,
  \doi{10.1126/science.207.4434.943}, 1980.

\bibitem[{\textit{Imbrie et~al.}(1992)}]{Imbrie92aa}
Imbrie, J., et~al., On the structure and origin of major glaciation cycles 1.
  {L}inear responses to {M}ilankovitch forcing, \textit{Paleoceanography},
  \textit{7}(6), 701--738, \doi{10.1029/93PA02751}, 1992.

\bibitem[{\textit{Imbrie et~al.}(1993)}]{imbrie93}
Imbrie, J., et~al., On the structure and origin of major glaciation cycles.
  {P}art 2: The 100, 000-year cycle, \textit{Paleoceanography}, \textit{8},
  699--735, 1993.

\bibitem[{\textit{Ionides et~al.}(2006)\textit{Ionides, Bret{\'o}, and
  King}}]{Ionides06aa}
Ionides, E.~L., C.~Bret{\'o}, and A.~A. King, Inference for nonlinear dynamical
  systems, \textit{Proceedings of the National Academy of Sciences},
  \textit{103}(49), 18,438--18,443, \doi{10.1073/pnas.0603181103}, 2006.

\bibitem[{\textit{Jaynes}(2003)}]{jaynes03aa}
Jaynes, E.~T., \textit{Probability Theory : The Logic of Science}, {Cambridge
  University Press}, Cambridge, UK, 2003.

\bibitem[{\textit{Kalnay}(2002)}]{Kalnay02aa}
Kalnay, E., \textit{Atmospheric Modeling, Data Assimilation and
  Predictability}, Cambrdige University Press, Cambridge, UK, 2002.

\bibitem[{\textit{Kaplan et~al.}(2003)}]{Kaplan03aa}
Kaplan, J.~O., et~al., Climate change and {A}rctic ecosystems: 2. {M}odeling,
  paleodata-model comparisons, and future projections, \textit{Journal of
  Geophysical Research}, \textit{108}(D19), \doi{10.1029/2002JD002559}, 2003.

\bibitem[{\textit{Kennedy and O'Hagan}(2000)}]{Kennedy00aa}
Kennedy, M.~C., and A.~O'Hagan, Predicting the output from a complex computer
  code when fast approximations are available, \textit{Biometrika},
  \textit{87}(1), 1--13, \doi{10.1093/biomet/87.1.1}, 2000.

\bibitem[{\textit{Kloeden and Platen}(1999)}]{Kloeden99aa}
Kloeden, P.~E., and E.~Platen, \textit{Numerical solution of stochastic
  differential equationts}, \textit{Application of mathematics, stochastic
  modelling and applied probability}, vol.~23, 3rd ed. ed., Springer Berlin /
  Heidelberg, 1999.

\bibitem[{\textit{K{\"o}hler et~al.}(2005)\textit{K{\"o}hler, Fischer,
  Munhoven, and Zeebe}}]{Kohler05aa}
K{\"o}hler, P., H.~Fischer, G.~Munhoven, and R.~E. Zeebe, Quantitative
  interpretation of atmospheric carbon records over the last glacial
  termination, \textit{Global Biogeochemical Cycles}, \textit{19}(4),
  \doi{10.1029/2004GB002345}, 2005.

\bibitem[{\textit{Korhola et~al.}(2002)\textit{Korhola, Vasko, Toivonen, and
  Olander}}]{Korhola02aa}
Korhola, A., K.~Vasko, H.~T.~T. Toivonen, and H.~Olander, Holocene temperature
  changes in northern {F}ennoscandia reconstructed from chironomids using
  {B}ayesian modelling, \textit{Quaternary Science Reviews},
  \textit{21}(16--17), 1841--1860, \doi{10.1016/S0277-3791(02)00003-3}, 2002.

\bibitem[{\textit{Kucera et~al.}(2005)\textit{Kucera, Rosell-Mel\'e, Schneider,
  Waelbroeck, and Winelt}}]{kucera05margo}
Kucera, M., A.~Rosell-Mel\'e, R.~Schneider, C.~Waelbroeck, and W.~Winelt,
  Multiproxy approach for the reconstruction of the glacial ocean surface
  ({M}{A}{R}{G}{O}), \textit{Quaternary Science Reviews}, \textit{24},
  813--819, \doi{10.1016/j.quascirev.2004.07.017}, 2005.

\bibitem[{\textit{Kutzbach and Liu}(1997)}]{Kutzbach97}
Kutzbach, J.~E., and Z.~Liu, Response of the {A}frican monsoon to orbital
  forcing and ocean feedbacks in the middle {H}olocene, \textit{Science},
  \textit{278}(5337), 440--443, \doi{10.1126/science.278.5337.440}, 1997.

\bibitem[{\textit{Kwasniok and Lohmann}(2009)}]{Kwasniok09aa}
Kwasniok, F., and G.~Lohmann, Deriving dynamical models from paleoclimatic
  records: Application to glacial millennial-scale climate variability,
  \textit{Phys. Rev. E}, \textit{80}, {066,104},
  \doi{10.1103/PhysRevE.80.066104}, 2009.

\bibitem[{\textit{LeGrande et~al.}(2006)\textit{LeGrande, Schmidt, Shindell,
  Field, Miller, Koch, Faluvegi, and Hoffman}}]{legrande06isotopes}
LeGrande, A.~N., G.~A. Schmidt, D.~T. Shindell, C.~V. Field, R.~L. Miller,
  D.~M. Koch, G.~Faluvegi, and G.~Hoffman, Consistent simulations of multiple
  proxy responses to an abrupt climate change event, \textit{Proceedings of the
  National Academy of Sciences}, \textit{103}, 837--842,
  \doi{10.1073/pnas.0510095103}, 2006.

\bibitem[{\textit{Lemieux-Dudon et~al.}(2010)\textit{Lemieux-Dudon, Blayo,
  Petit, Waelbroeck, Svensson, Ritz, Barnola, Narcisi, and
  Parrenin}}]{Lemieux-Dudon10aa}
Lemieux-Dudon, B., E.~Blayo, J.-R. Petit, C.~Waelbroeck, A.~Svensson, C.~Ritz,
  J.-M. Barnola, B.~M. Narcisi, and F.~Parrenin, Consistent dating for
  {A}ntarctic and {G}reenland ice cores, \textit{Quaternary Science Reviews},
  \textit{29}(1--2), 8--20, \doi{10.1016/j.quascirev.2009.11.010}, 2010.

\bibitem[{\textit{Li et~al.}(2010)\textit{Li, Nychka, and Ammann}}]{Li10aa}
Li, B., D.~W. Nychka, and C.~M. Ammann, The value of multiproxy reconstruction
  of past climate, \textit{Journal of the American Statistical Association},
  \textit{105}(491), 883--895, \doi{10.1198/jasa.2010.ap09379}, 2010.

\bibitem[{\textit{Linvill}(1953)}]{Linvill53aa}
Linvill, J.~G., Transistor negative-impedence converters, \textit{Proceedings
  of the IRE}, \textit{41}, 725--729, 1953.

\bibitem[{\textit{Lisiecki and Raymo}(2007)}]{lisiecki07trends}
Lisiecki, L.~E., and M.~E. Raymo, Plio-{P}leistocene climate evolution: trends
  and transitions in glacial cycles dynamics, \textit{Quaternary Sci. Rev.},
  \textit{26}, 56--69, \doi{10.1016/j.quascirev.2006.09.005}, 2007.

\bibitem[{\textit{Lisiecki et~al.}(2008)\textit{Lisiecki, Raymo, and
  Curry}}]{Lisiecki08ab}
Lisiecki, L.~E., M.~E. Raymo, and W.~B. Curry, Atlantic overturning responses
  to {L}ate {P}leistocene climate forcings, \textit{Nature},
  \textit{456}(7218), 85--88, \doi{10.1038/nature07425}, 2008.

\bibitem[{\textit{Liu and West}(2001)}]{lw01}
Liu, J., and M.~West, Combined parameter and state estimation in
  simulation-based filtering, in \textit{Sequential {M}onte {C}arlo Methods in
  Practice}, edited by A.~Doucet, N.~de~Freitas, and N.~Gordon, Springer, 2001.

\bibitem[{\textit{Livina et~al.}(2011)\textit{Livina, Kwasniok, Lohmann,
  Kantelhardt, and Lenton}}]{Livina11aa}
Livina, V.~N., F.~Kwasniok, G.~Lohmann, J.~W. Kantelhardt, and T.~M. Lenton,
  Changing climate states and stability: from pliocene to present,
  \textit{Climate Dynamics}, \textit{37}(11-12), 2437--2453,
  \doi{10.1007/s00382-010-0980-2}, 2011.

\bibitem[{\textit{Ljung}(1999)}]{Ljung99aa}
Ljung, L., \textit{System identification -- Theory for the User}, Prentice
  Hall, Upper-Saddle River, N. J., 1999.

\bibitem[{\textit{Majda et~al.}(2009)\textit{Majda, Franzke, and
  Crommelin}}]{Majda09aa}
Majda, A.~J., C.~Franzke, and D.~Crommelin, Normal forms for reduced stochastic
  climate models, \textit{Proceedings of the National Academy of Sciences of
  the United States of America}, \textit{106}(10), 3649--3653,
  \doi{10.1073/pnas.0900173106}, 2009.

\bibitem[{\textit{Marchal et~al.}(2000)\textit{Marchal, Francois, Stocker, and
  Joos}}]{Marchal2000path}
Marchal, O., R.~Francois, T.~Stocker, and F.~Joos, Ocean thermohaline
  circulation and sedimentary $^{231}${P}a/$^{230}${T}h ratio,
  \textit{Paleoceanography}, \textit{15}, 625--641, \doi{10.1029/2000PA000496},
  2000.

\bibitem[{\textit{{MARGO project members}}(2009)}]{margo09aa}
{MARGO project members}, Constraints on the magnitude and patterns of ocean
  cooling at the {L}ast {G}lacial {M}aximum, \textit{Nature Geosciences},
  \textit{2}(2), 127--132, \doi{10.1038/ngeo411}, 2009.

\bibitem[{\textit{Montoya et~al.}(2005)\textit{Montoya, Griesel, Levermann,
  Mignot, Hofmann, Ganopolski, and Rahmstorf}}]{Montoya05aa}
Montoya, M., A.~Griesel, A.~Levermann, J.~Mignot, M.~Hofmann, A.~Ganopolski,
  and S.~Rahmstorf, The earth system model of intermediate complexity
  {CLIMBER-3$\alpha$}. {P}art {I}: description and performance for present-day
  conditions, \textit{Climate Dynamics}, \textit{25}(2), 237--263,
  \doi{10.1007/s00382-005-0044-1}, 2005.

\bibitem[{\textit{Mouchet and Deleersnijder}(2008)}]{Mouchet08aa}
Mouchet, A., and E.~Deleersnijder, The leaky funnel model, a metaphor of the
  ventilation of the {W}orld {O}cean as simulated in an {OGCM}, \textit{Tellus
  A}, \textit{60}(4), 761--774, \doi{10.1111/j.1600-0870.2008.00322.x}, 2008.

\bibitem[{\textit{Mudelsee}(2000)}]{Manfred00aa}
Mudelsee, M., Ramp function regression: a tool for quantifying climate
  transitions, \textit{Computers \& Geosciences}, \textit{26}(3), 293--307,
  \doi{10.1016/S0098-3004(99)00141-7}, 2000.

\bibitem[{\textit{Mudelsee and Raymo}(2005)}]{Mudelsee05aa}
Mudelsee, M., and M.~E. Raymo, Slow dynamics of the {N}orthern {H}emisphere
  glaciation, \textit{Paleoceanography}, \textit{20}, PA4022,
  \doi{10.1029/2005PA001153}, 2005.

\bibitem[{\textit{M{\"u}ller et~al.}(2006)\textit{M{\"u}ller, Joos, Edwards,
  and Stocker}}]{Muller06aa}
M{\"u}ller, S.~A., F.~Joos, N.~R. Edwards, and T.~F. Stocker, Water mass
  distribution and ventilation time scales in a cost-efficient,
  three-dimensional ocean model, \textit{Journal of Climate}, \textit{19}(21),
  5479--5499, \doi{10.1175/JCLI3911.1}, 2006.

\bibitem[{\textit{Munhoven}(2007)}]{Munhoven07aa}
Munhoven, G., Glacial-interglacial rain ratio changes: Implications for
  atmospheric {CO$_2$} and ocean-sediment interaction, \textit{Deep Sea
  Research Part II: Topical Studies in Oceanography}, \textit{54}(5-7),
  722--746, \doi{10.1016/j.dsr2.2007.01.008}, 2007.

\bibitem[{\textit{Oakley and O'Hagan}(2002)}]{Oakley02aa}
Oakley, J., and A.~O'Hagan, Bayesian inference for the uncertainty distribution
  of computer model outputs, \textit{Biometrika}, \textit{89}(4), 769--784,
  \doi{10.1093/biomet/89.4.769}, 2002.

\bibitem[{\textit{Oakley and O'Hagan}(2004)}]{Oakley04aa}
Oakley, J.~E., and A.~O'Hagan, Probabilistic sensitivity analysis of complex
  models: a {B}ayesian approach, \textit{Journal of the Royal Statistical
  Society: Series B (Statistical Methodology)}, \textit{66}(3), 751--769,
  \doi{10.1111/j.1467-9868.2004.05304.x}, 2004.

\bibitem[{\textit{Oerlemans}(1980)}]{oerlemans80}
Oerlemans, J., Model experiments on the 100,000-yr glacial cycle,
  \textit{Nature}, \textit{287}(2), 430--432,
  \doi{http://dx.doi.org/10.1038/287430a0}, 1980.

\bibitem[{\textit{Oerlemans}(1981)}]{Oerlemans81aa}
Oerlemans, J., Some basic experiments with a vertically-integrated ice sheet
  model, \textit{Tellus}, \textit{33}(1), 1--11,
  \doi{10.1111/j.2153-3490.1981.tb01726.x}, 1981.

\bibitem[{\textit{Olson et~al.}(2012)\textit{Olson, Sriver, Goes, Urban,
  Matthews, Haran, and Keller}}]{Olson12aa}
Olson, R., R.~Sriver, M.~Goes, N.~M. Urban, H.~D. Matthews, M.~Haran, and
  K.~Keller, A climate sensitivity estimate using {B}ayesian fusion of
  instrumental observations and an {E}arth system model, \textit{Journal of
  Geophysical Research}, \textit{117}(D4), D04,103, \doi{10.1029/2011JD016620},
  2012.

\bibitem[{\textit{Oreskes et~al.}(1994)\textit{Oreskes, Shrader-Frechette, and
  Belitz}}]{Oreskes04021994}
Oreskes, N., K.~Shrader-Frechette, and K.~Belitz, Verification, validation, and
  confirmation of numerical models in the earth sciences, \textit{Science},
  \textit{263}(5147), 641--646, \doi{10.1126/science.263.5147.641}, 1994.

\bibitem[{\textit{Otto-Bliesner et~al.}(2009)}]{Otto-Bliesner09aa}
Otto-Bliesner, B.~L., et~al., A comparison of {PMIP2} model simulations and the
  {MARGO} proxy reconstruction for tropical sea surface temperatures at last
  glacial maximum, \textit{Climate Dynamics}, \textit{32}(6), 799--815,
  \doi{10.1007/s00382-008-0509-0}, 2009.

\bibitem[{\textit{Paillard}(1995)}]{Paillard95aa}
Paillard, D., The hierarchical structure of glacial climatic oscillations:
  interactions between ice-sheet dynamics and climate, \textit{Climate
  Dynamics}, \textit{11}, 162--177, \doi{10.1007/BF00223499}, 1995.

\bibitem[{\textit{Paillard}(1998)}]{paillard98}
Paillard, D., The timing of {P}leistocene glaciations from a simple
  multiple-state climate model, \textit{Nature}, \textit{391}, 378--381,
  \doi{10.1038/34891}, 1998.

\bibitem[{\textit{Paillard}(2001)}]{paillard01rge}
Paillard, D., Glacial cycles: Toward a new paradigm, \textit{Reviews of
  Geophysics}, \textit{39}(3), 325--346, \doi{10.1029/2000RG00009}, 2001.

\bibitem[{\textit{Paillard and Labeyrie}(1994)}]{paillard94aa}
Paillard, D., and L.~Labeyrie, Role of the thermohaline circulation in the
  abrupt warming after {H}einrich events, \textit{Nature}, \textit{372}(6502),
  162--164, \doi{10.1038/372162a0}, 1994.

\bibitem[{\textit{Paillard and Parrenin}(2004)}]{paillard04eps}
Paillard, D., and F.~Parrenin, The {A}ntarctic ice sheet and the triggering of
  deglaciations, \textit{Earth Planet. Sc. Lett.}, \textit{227}, 263--271,
  \doi{10.1016/j.epsl.2004.08.023}, 2004.

\bibitem[{\textit{Palmer}(2005)}]{Palmer05aa}
Palmer, T., Global warming in a nonlinear climate - can we be sure?,
  \textit{Europhysics News}, \textit{36}(2), 42--46, \doi{10.1051/epn:2005202},
  2005.

\bibitem[{\textit{Paul and Losch}(2012)}]{Paul12aa}
Paul, A., and M.~Losch, Perspectives of parameter and state estimation in
  palaeoclimatology, in \textit{Climate Change: Inferences from Paleoclimate
  and Regional Aspects}, edited by M.~F. Berger, A. and D.~Sijakci, pp.
  93--105, \doi{10.1007/978-3-7091-0973-1\_7}, 2012.

\bibitem[{\textit{Paul and Sch{\"a}fer-Neth}(2005)}]{Paul05aa}
Paul, A., and C.~Sch{\"a}fer-Neth, How to combine sparse proxy data and coupled
  climate models, \textit{Quaternary Science Reviews}, \textit{24}(7--9),
  1095--1107, \doi{10.1016/j.quascirev.2004.05.010}, 2005.

\bibitem[{\textit{Penland}(2007)}]{Penland07aa}
Penland, C., Stochastic linear models of nonlinear geosystems, in
  \textit{Nonlinear Dynamics in Geosciences}, edited by A.~A. Tsonis and J.~B.
  Elsner, pp. 485--515, Springer New York, 2007.

\bibitem[{\textit{Petoukhov et~al.}(2000)\textit{Petoukhov, Ganopolski,
  Brovkin, Claussen, Eliseev, Kubatzki, and Rahmstorf}}]{petoukhov00}
Petoukhov, V., A.~Ganopolski, V.~Brovkin, M.~Claussen, A.~Eliseev, C.~Kubatzki,
  and S.~Rahmstorf, {CLIMBER-2}: a climate system model of intermediate
  complexity. part {I}: model description and performance for present climate,
  \textit{Climate Dynamics}, \textit{16}, 1--17, \doi{10.1007/PL00007919`},
  2000.

\bibitem[{\textit{Pikovski et~al.}(2001)\textit{Pikovski, Rosenblum, and
  Kurths}}]{Pikovski01aa}
Pikovski, A., M.~Rosenblum, and J.~Kurths, \textit{Synchronization: a universal
  concept in nonlinear sciences}, \textit{Cambridge Nonlinear Science Series},
  vol.~12, Cambrdige University Press, New York, 2001.

\bibitem[{\textit{Pollard}(1983)}]{Pollard83aa}
Pollard, D., A coupled climate-ice sheet model applied to the {Q}uaternary ice
  ages, \textit{Journal of Geophysical Research}, \textit{88}(C12), 7705--7718,
  \doi{10.1029/JC088iC12p07705}, 1983.

\bibitem[{\textit{Rahmstorf}(2000)}]{rahmstorf00thresholds}
Rahmstorf, S., The thermohaline circulation: a system with dangerous
  thresholds?, \textit{Climatic Change}, \textit{46}, 247--256,
  \doi{10.1023/A:1005648404783}, 2000.

\bibitem[{\textit{Rasmussen and Williams}(2005)}]{Rasmussen06aa}
Rasmussen, C., and C.~Williams, \textit{Gaussian Processes for Machine
  Learning}, Adaptive Computation And Machine Learning, MIT Press, Cambridge
  MA, 2005.

\bibitem[{\textit{Renssen et~al.}(2009)\textit{Renssen, Seppa, Heiri, Roche,
  Goosse, and Fichefet}}]{Renssen09aa}
Renssen, H., H.~Seppa, O.~Heiri, D.~M. Roche, H.~Goosse, and T.~Fichefet, The
  spatial and temporal complexity of the {H}olocene thermal maximum,
  \textit{Nature Geosciences}, \textit{2}(6), 411--414, \doi{10.1038/ngeo513},
  2009.

\bibitem[{\textit{Ridgwell et~al.}(2007)\textit{Ridgwell, Hargreaves, Edwards,
  Annan, Lenton, Marsh, Yool, and Watson}}]{Ridgwell07aa}
Ridgwell, A., J.~C. Hargreaves, N.~R. Edwards, J.~D. Annan, T.~M. Lenton,
  R.~Marsh, A.~Yool, and A.~Watson, Marine geochemical data assimilation in an
  efficient earth system model of global biogeochemical cycling,
  \textit{Biogeosciences}, \textit{4}(1), 87--104, \doi{10.5194/bg-4-87-2007},
  2007.

\bibitem[{\textit{Ritz et~al.}(2010)\textit{Ritz, Stocker, and
  Joos}}]{Ritz10ab}
Ritz, S.~P., T.~F. Stocker, and F.~Joos, A coupled dynamical ocean--energy
  balance atmosphere model for paleoclimate studies, \textit{Journal of
  Climate}, \textit{24}(2), 349--375, \doi{10.1175/2010JCLI3351.1}, 2010.

\bibitem[{\textit{Roche et~al.}(2004)\textit{Roche, Paillard, and
  Cortijo}}]{Roche04aa}
Roche, D., D.~Paillard, and E.~Cortijo, Constraints on the duration and
  freshwater release of {H}einrich event 4 through isotope modelling,
  \textit{Nature}, \textit{432}(7015), 379--382, \doi{10.1038/nature03059},
  2004.

\bibitem[{\textit{Roche et~al.}(2006)\textit{Roche, Donnadieu, Puc{\'e}at, and
  Paillard}}]{Roche06aa}
Roche, D.~M., Y.~Donnadieu, E.~Puc{\'e}at, and D.~Paillard, Effect of changes
  in $\delta^{18}${O} content of the surface ocean on estimated sea surface
  temperatures in past warm climate, \textit{Paleoceanography}, \textit{21}(2),
  PA2023, \doi{10.1029/2005PA001220}, 2006.

\bibitem[{\textit{Rougier}(2007)}]{Rougier2007Probabilistic-i}
Rougier, J., Probabilistic inference for future climate using an ensemble of
  climate model evaluations, \textit{Climatic Change}, \textit{81}, 247--264,
  \doi{10.1007/s10584-006-9156-9}, 2007.

\bibitem[{\textit{Rougier}(2008)}]{Rougier08aa}
Rougier, J., Efficient emulators for multivariate deterministic functions,
  \textit{Journal of Computational and Graphical Statistics}, \textit{17}(4),
  827--843, \doi{10.1198/106186008X384032}, 2008.

\bibitem[{\textit{Ruddiman}(2006)}]{Ruddiman06aa}
Ruddiman, W.~F., Ice-driven {CO}$_2$ feedback on ice volume, \textit{Climate of
  the Past}, \textit{2}, 43--55, \doi{10.5194/cp-2-43-2006}, 2006.

\bibitem[{\textit{Saltzman}(1962)}]{Saltzman62aa}
Saltzman, B., Finite amplitude free convection as an initial value
  problem---{I}, \textit{Journal of the Atmospheric Sciences}, \textit{19}(4),
  329--341, \doi{10.1175/1520-0469(1962)019<0329:FAFCAA>2.0.CO;2}, 1962.

\bibitem[{\textit{Saltzman}(2001)}]{saltzman02book}
Saltzman, B., \textit{Dynamical paleoclimatology: Generalized Theory of Global
  Climate Change (International Geophysics)}, \textit{International Geophysics
  Series}, vol.~80, Academic {P}ress, 2001.

\bibitem[{\textit{Saltzman and Maasch}(1990)}]{saltzman90sm}
Saltzman, B., and K.~A. Maasch, A first-order global model of late {C}enozoic
  climate, \textit{Transactions of the Royal Society of Edinburgh Earth
  Sciences}, \textit{81}, 315--325, \doi{10.1017/S0263593300020824}, 1990.

\bibitem[{\textit{Santner et~al.}(2003)\textit{Santner, Williams, and
  Notz}}]{santner03}
Santner, T., B.~Williams, and W.~Notz, \textit{The Design and Analysis of
  Computer Experiments}, New York: Springer, 2003.

\bibitem[{\textit{Sch{\"a}fer-Neth et~al.}(2005)\textit{Sch{\"a}fer-Neth, Paul,
  and Mulitza}}]{Schafer-Neth05aa}
Sch{\"a}fer-Neth, C., A.~Paul, and S.~Mulitza, Perspectives on mapping the
  {MARGO} reconstructions by variogram analysis/kriging and objective analysis,
  \textit{Quaternary Science Reviews}, \textit{24}(7--9), 1083--1093,
  \doi{10.1016/j.quascirev.2004.06.017}, 2005.

\bibitem[{\textit{Scheffer et~al.}(2009)}]{Scheffer09ab}
Scheffer, M., et~al., Early-warning signals for critical transitions,
  \textit{Nature}, \textit{461}(7260), 53--59, \doi{10.1038/nature08227}, 2009.

\bibitem[{\textit{Schmittner et~al.}(2011)\textit{Schmittner, Urban, Shakun,
  Mahowald, Clark, Bartlein, Mix, and Rosell-Mel{\'e}}}]{Schmittner11aa}
Schmittner, A., N.~M. Urban, J.~D. Shakun, N.~M. Mahowald, P.~U. Clark, P.~J.
  Bartlein, A.~C. Mix, and A.~Rosell-Mel{\'e}, Climate sensitivity estimated
  from temperature reconstructions of the {L}ast {G}lacial {M}aximum,
  \textit{Science}, \textit{334}(6061), 1385--1388,
  \doi{10.1126/science.1203513}, 2011.

\bibitem[{\textit{Schneider~von Deimling et~al.}(2006)\textit{Schneider~von
  Deimling, Held, Ganopolski, and Rahmstorf}}]{schneidervd06lgm}
Schneider~von Deimling, T., H.~Held, A.~Ganopolski, and S.~Rahmstorf, Climate
  sensitivity estimated from ensemble simulations of glacial climate,
  \textit{Climate Dynamics}, \textit{27}, 149--163,
  \doi{10.1007/s00382-006-0126-8}, 2006.

\bibitem[{\textit{Schulz et~al.}(2002)\textit{Schulz, Paul, and
  Timmermann}}]{Schulz2002oscillators}
Schulz, M., A.~Paul, and A.~Timmermann, Relaxation oscillators in concert: a
  framework for climate change at millennial timescales during the late
  {P}leistocene, \textit{Geophysical Research Letters}, \textit{29}(24), 2193,
  \doi{10.1029/2002GL016144}, 2002.

\bibitem[{\textit{Schwartz}(2011)}]{Schwartz11aa}
Schwartz, S., Feedback and sensitivity in an electrical circuit: an analog for
  climate models, \textit{Climatic Change}, \textit{106}(2), 315--326,
  \doi{10.1007/s10584-010-9903-9}, 2011.

\bibitem[{\textit{Shackleton}(2000)}]{shackleton00}
Shackleton, N.~J., The 100,000-year ice-age cycle identified and found to lag
  temperature, carbon dioxide and orbital eccentricity, \textit{Science},
  \textit{289}, 1897--1902, \doi{10.1126/science.289.5486.1897}, 2000.

\bibitem[{\textit{Smith}(2012)}]{Smith12aa}
Smith, R.~S., The {FAMOUS} climate model (versions {XFXWB} and {XHCC}):
  description update to version {XDBUA}, \textit{Geoscientific Model
  Development}, \textit{5}(1), 269--276, \doi{10.5194/gmd-5-269-2012}, 2012.

\bibitem[{\textit{Stommel}(1961)}]{stommel61}
Stommel, H., Thermohaline convection with two stable regimes of flow,
  \textit{Tellus}, \textit{13}, 224--230,
  \doi{10.1111/j.2153-3490.1961.tb00079.x}, 1961.

\bibitem[{\textit{Timmermann et~al.}(2003)\textit{Timmermann, Gildor, Schulz,
  and Tziperman}}]{Timmermann03ab}
Timmermann, A., H.~Gildor, M.~Schulz, and E.~Tziperman, Coherent resonant
  millennial-scale climate oscillations triggered by massive meltwater pulses,
  \textit{Journal of Climate}, \textit{16}(15), 2569--2585,
  \doi{10.1175/1520-0442(2003)016<2569:CRMCOT>2.0.CO;2}, 2003.

\bibitem[{\textit{Tingley and Huybers}(2010)}]{Tingley10aa}
Tingley, M.~P., and P.~Huybers, A bayesian algorithm for reconstructing climate
  anomalies in space and time. {P}art {I}: Development and applications to
  paleoclimate reconstruction problems., \textit{Journal of Climate},
  \textit{23}(10), 2759--2781, \doi{10.1175/2009JCLI3015.1}, 2010.

\bibitem[{\textit{Tingley et~al.}(2012)\textit{Tingley, Craigmile, Haran, Li,
  Mannshardt, and Rajaratnam}}]{Tingley12aa}
Tingley, M.~P., P.~F. Craigmile, M.~Haran, B.~Li, E.~Mannshardt, and
  B.~Rajaratnam, Piecing together the past: statistical insights into
  paleoclimatic reconstructions, \textit{Quaternary Science Reviews},
  \textit{35}, 1--22, \doi{10.1016/j.quascirev.2012.01.012}, 2012.

\bibitem[{\textit{Tziperman et~al.}(1994)\textit{Tziperman, Stone, Cane, and
  Jarosh}}]{tziperman94}
Tziperman, E., L.~Stone, M.~A. Cane, and H.~Jarosh, {El-Ni\~{n}o} chaos:
  {O}verlapping of resonances between the seasonal cycle and the pacific
  ocean-atmosphere oscillator, \textit{Science}, \textit{264}, 72--74,
  \doi{10.1126/science.264.5155.72}, 1994.

\bibitem[{\textit{Tziperman et~al.}(2006)\textit{Tziperman, Raymo, Huybers, and
  Wunsch}}]{tziperman06pacing}
Tziperman, E., M.~E. Raymo, P.~Huybers, and C.~Wunsch, Consequences of pacing
  the {P}leistocene 100 kyr ice ages by nonlinear phase locking to
  {M}ilankovitch forcing, \textit{Paleoceanography}, \textit{21}, PA4206,
  \doi{10.1029/2005PA001241}, 2006.

\bibitem[{\textit{Urban and Fricker}(2010)}]{Urban10aa}
Urban, N.~M., and T.~E. Fricker, A comparison of latin hypercube and grid
  ensemble designs for the multivariate emulation of an earth system model,
  \textit{Computers \& Geosciences}, \textit{36}(6), 746--755,
  \doi{10.1016/j.cageo.2009.11.004}, 2010.

\bibitem[{\textit{Voss et~al.}(2004)\textit{Voss, Timmer, and
  Kurths}}]{Voss04aa}
Voss, H.~U., J.~Timmer, and J.~Kurths, Nonlinear dynamical system
  identification form uncertain and indirect measurements,
  \textit{International Journal of Bifurcation and Chaos}, \textit{14}(6),
  1905--1933, \doi{10.1142/S0218127404010345}, 2004.

\bibitem[{\textit{Wilkinson}(2010)}]{Wilkinson10aa}
Wilkinson, R.~D., Bayesian calibration of expensive multivariate computer
  experiments, in \textit{Large-Scale Inverse Problems and Quantification of
  Uncertainty}, edited by L.~Biegler, G.~Biros, O.~Ghattas, M.~Heinkenschloss,
  D.~Keyes, B.~Mallick, Y.~Marzouk, L.~Tenorio, B.~van Bloemen~Waanders, and
  K.~Willcox, pp. 195--215, John Wiley \& Sons, Ltd, Chichester, UK.,
  \doi{10.1002/9780470685853.ch10}, 2010.

\bibitem[{\textit{Wilson et~al.}(2008)\textit{Wilson, Bushell, Kerr-Munslow,
  Price, Morcrette, and Bodas-Salcedo}}]{Wilson08aa}
Wilson, D.~R., A.~C. Bushell, A.~M. Kerr-Munslow, J.~D. Price, C.~J. Morcrette,
  and A.~Bodas-Salcedo, {PC}2: A prognostic cloud fraction and condensation
  scheme. ii: Climate model simulations, \textit{Quarterly Journal of the Royal
  Meteorological Society}, \textit{134}(637), 2109--2125, \doi{10.1002/qj.332},
  2008.

\bibitem[{\textit{Winsberg}(1999)}]{Winsberg99aa}
Winsberg, E., Sanctioning models: The epistemology of simulation,
  \textit{Science in Context}, \textit{12}(02), 275--292,
  \doi{10.1017/S0269889700003422}, 1999.

\bibitem[{\textit{Winton and Sarachik}(1993)}]{Winton93aa}
Winton, M., and E.~S. Sarachik, Thermohaline oscillations induced by strong
  steady salinity forcing of ocean general circulation models, \textit{Journal
  of Physical Oceanography}, \textit{23}(7), 1389--1410,
  \doi{10.1175/1520-0485(1993)023<1389:TOIBSS>2.0.CO;2}, 1993.

\bibitem[{\textit{Wunsch}(2006)}]{Wunsch06aa}
Wunsch, C., \textit{Discrete Inverse and State Estimation Problems: With
  Geophysical Fluid Applications}, Cambrdige University Press, Cambridge, UK,
  2006.

\bibitem[{\textit{Yin et~al.}(2008)\textit{Yin, Berger, Driesschaert, Goosse,
  Loutre, and Crucifix}}]{yin08}
Yin, Q., A.~Berger, E.~Driesschaert, H.~Goosse, M.~F. Loutre, and M.~Crucifix,
  The {E}urasian ice sheet reinforces the east asian summer monsoon during the
  interglacial 500 000 years ago, \textit{Climate of the Past}, \textit{4}(2),
  79--90, \doi{10.5194/cp-4-79-2008}, 2008.

\bibitem[{\textit{Young and Ratto}(2009)}]{Young09aa}
Young, P., and M.~Ratto, A unified approach to environmental systems modeling,
  \textit{Stochastic Environmental Research and Risk Assessment},
  \textit{23}(7), 1037--1057, \doi{10.1007/s00477-008-0271-1}, 2009.

\bibitem[{\textit{Young and Ratto}(2011)}]{Young11aa}
Young, P.~C., and M.~Ratto, Statistical emulation of large linear dynamic
  models, \textit{Technometrics}, \textit{53}(1), 29--43,
  \doi{10.1198/TECH.2010.07151}, 2011.

\bibitem[{\textit{Zhao et~al.}(2005)}]{zhao05mh}
Zhao, Y., et~al., A multi-model analysis of the role of the ocean on the
  {A}frican and {I}ndian monsoon during the mid-{H}olocene, \textit{Climate
  Dynamics}, \textit{25}, 777--800, \doi{10.1007/s00382-005-0075-7}, 2005.

\end{thebibliography}
\end{document}